\documentclass[aps,prb,reprint,groupedaddress,citeautoscript,showpacs]{revtex4-1}

\usepackage{srcltx}
\usepackage{graphicx}
\usepackage{epstopdf}

\usepackage{amsmath} 
\usepackage{amsfonts}
\usepackage{amssymb}
\DeclareMathOperator{\Tr}{Tr}

\newcommand{\sminus}{-}

\renewcommand{\vec}[1]{\mathbf{#1}}

\begin{document}

\title{Quantum theory of multimode polariton condensation}
\author{David Racine}
\author{P. R. Eastham}

\affiliation{School of Physics and CRANN, Trinity College Dublin, Dublin 2, Ireland}

\date{\today}

\begin{abstract} 
  We develop a theory for the dynamics of the density matrix
  describing a multimode polariton condensate. In such a condensate
  several single-particle orbitals become highly occupied, due to
  stimulated scattering from reservoirs of high-energy excitons. A
  generic few-parameter model for the system leads to a Lindblad
  equation which includes saturable pumping, decay, and condensate
  interactions.  We show how this theory can be used to obtain the
  population distributions, and the time-dependent first- and
  second-order coherence functions, in such a multimode condensate. As
  a specific application, we consider a polaritonic Josephson
  junction, formed from a double-well potential. We obtain the
  population distributions, emission line shapes, and widths
  (first-order coherence functions), and predict the dephasing time of
  the Josephson oscillations. \end{abstract}

\maketitle

\section{Introduction}

The strong coupling of quantum well excitons and microcavity photons
gives rise to part-light and part-matter quasiparticles known as
cavity polaritons.\cite{Hopfield1958,Weisbuch1992} Polaritons inherit
the bosonic nature of their constituents, allowing them to undergo
Bose-Einstein condensation.\cite{Dang1998,Kasprzak2006,Love2008,AssmannForchel2011} The
condensation is characterized by the appearance of macroscopically occupied single-particle 
states in a pumped
microcavity. It differs from Bose-Einstein condensation in atomic
gases,\cite{Pitaevskii2003} because the polaritons have a very light
effective mass, and can therefore condense at a much higher
background temperature. Another key distinction between polariton condensates and
their atomic counterparts is the fact that polaritons decay into
external photons, typically in a few picoseconds. Thus,
the polariton condensate is a nonequilibrium steady-state, maintained
by a balance between the radiative decay and the external pumping. The external pumping 
generally creates a population of excitons or polaritons at high energy, and condensation occurs in 
lower energy states due to stimulated scattering from the high-energy reservoir.

One consequence of this nonequilibrium nature is that, whereas in
equilibrium only the lowest energy single-particle state can be
macroscopically occupied, for polaritons large occupations can build
up in other orbitals. Furthermore, the condensation can occur in
several orbitals simultaneously, enabling the study of interacting
macroscopic quantum states. The presence of several highly occupied
states of a trapping potential can be seen directly in the emission
spectra,\cite{TosiBaumberg2012NatPhys,GalbiatiBloch2012} and inferred
from the presence of Josephson oscillations.\cite{Lagoudakis2010, BlochJoseph2013} Such
multimode condensation can also occur in spatially extended states, in
particular in the Bloch states of one\cite{LaiYamamoto2007Nat} and
two\cite{MasumotoYamamoto2012Bloch} dimensional lattices. The in-plane
potentials that control these condensates can arise from
growth induced disorder in the Bragg
mirrors,\cite{Langbein1999,Love2008,Lagoudakis2010} metal-film
patterning of the mirror
surfaces,\cite{LaiYamamoto2007Nat,MasumotoYamamoto2012Bloch} and
interaction effects,\cite {TosiBaumberg2012NatPhys} as well as from
the use of non-planar structures such as micropillars and photonic
molecules.\cite{GalbiatiBloch2012}

The theoretical modeling of these nonequilibrium quantum objects has
been performed quite exten\-si\-vely within the mean-field approximation,
using an augmented Gross-Pitaevskii description\cite{WoutersCarusoto2007,Keeling2008,Eastham2008, Rodrigues2012} to
treat the dynamics of the highly occupied orbitals, and to obtain the
excitation spectra. Mean-field solutions of microscopic models using
the nonequilibrium Green's function
formalism\cite{SzymanskaKeelingLittlewood2007} have also been
developed. The Langevin\cite{TassoneYamamoto2000,HaugThoai2012, wouters09},
Fokker-Planck\cite{Whittaker2009,SchwendimannQuattropaniSarchi2010},
and density
matrix\cite{PorrasTejedor2003,LaussyBigenwald2004,VergerCarusotto2006,SchendimannQuattropani2008,Shelykh2011}
frameworks have been used to derive the quantum statistics of the
condensate, and hence the first- and second-order coherence functions
of the optical emission.  The density matrix approach allows the treatment of fluctuations in the condensate as well as the direct incorporation of incoherent phenomena such as the interaction with phonon baths.\cite{LaussyBigenwald2004,Shelykh2011} While mean field theories are already able
to treat several highly occupied
orbitals,\cite{Eastham2008,Wouters2008,Rodrigues2012} full quantum
treatments of this regime have yet to be formulated.

The aim of this work is to develop a density matrix approach for
multimode polariton condensation, in which several single-particle
orbitals are driven by several reservoirs. It treats both quantum and
nonequilibrium fluctuations, and allows photon statistics and emission
spectra to be calculated. We first derive a Lindblad equation for two
condensate modes (highly occupied orbitals) pumped by a single
reservoir of higher energy particles, using a treatment similar to
that of a two-mode laser,\cite{SinghZubairy1980} and then extend the
result to treat several reservoirs pumping several condensates. This
gives a generic model for the quantum dynamics of a nonequilibrium
polariton condensate, with the complexity of the reservoirs captured
in a few known parameters. We show how the theory can be used to
obtain the population distribution of the condensate orbitals, both
numerically and analytically. We also show how it may be used to
calculate both first-order and second-order coherence functions. The
first-order coherence function, $\langle a_1^\dagger(\tau) a_1
\rangle$, is the Fourier transform of the emission spectrum from one
condensate orbital. We obtain it in three different ways: (i) direct
numerical solution of the Lindblad form, (ii) making a continuum
approximation to obtain a soluble partial differential equation,
related to the Fokker-Planck equation,\cite{VanKampen,Whittaker2009}
and (iii) a cruder static limit approximation, which neglects the
dynamics of the populations, but is generally valid near
threshold.\cite{Whittaker2009} Among higher order correlation
functions we consider those of the form $\langle a_1^\dagger(\tau)
a_2(\tau) a_2^\dagger a_1 \rangle$, which quantify the dephasing of
the intensity oscillations that correspond to the beating between the
condensate modes. Such oscillations are a form of Josephson
oscillations, which have been observed
experimentally.\cite{Lagoudakis2010} We analyze their dephasing both
numerically and in the static limit.

As a specific application of our theory, we study fluctuations in a
polariton Josephson junction, formed in a double-well
potential,\cite{Lagoudakis2010} using a tight-binding model in which
each well is pumped by a corresponding reservoir. Diagonalizing the
Hamiltonian leads to symmetric and antisymmetric orbitals when the
wells are degenerate. We obtain the population distribution in these
orbitals, calculate the emission line shapes and widths, and predict
the dephasing time of the Josephson oscillations in the quasilinear
regime, where interactions have a negligible effect on the mean-field
dynamics. We predict large fluctuations in the populations when the
wells are tuned to resonance, due to the presence of a soft density
mode, and show how the emission is broadened by intermode and
intramode interactions.

The remainder of this paper is structured as follows. In
Sec. \ref{sec:pumpmodel}, we give the Lindblad form for the pumping of
two condensate orbitals by one reservoir [Eq. (\ref{eq:4})], and
obtain the generalization to many condensate orbitals pumped by many
reservoirs [Eq. (\ref{eq:8})]. We also provide expressions for the
population distributions. In Sec. \ref{sec:coherencefun} we introduce
an approximate form for the Hamiltonian dynamics of the condensates,
and show how coherence functions can be
obtained. Sec. \ref{sec:twomodes} addresses the specific problem of
the double-well potential, giving results for the population
distributions (Fig. \ref{fig:populations}), the decay of
first-order-coherence (Fig. \ref{fig:linewidth}), the variation in
coherence time (Fig. \ref{fig:tc_effect}), and the dephasing of
intensity (Josephson) oscillations (Fig. \ref{fig:Joseph}). Finally,
in Secs. \ref{sec:discussion} and \ref{sec:conclusion} we discuss wider
applications of our results, summarize our conclusions, and outline
some suggestions for future work.

\section{\label{sec:pumpmodel} The pumping model}

In this section, we develop a model for the
pumping of low-energy polariton orbitals by scattering from a
high-energy reservoir of excitons or polaritons. We consider first two
condensates being pumped by a single reservoir, and then generalize
the result to many condensates pumped by many reservoirs. Our final
aim is an equation of motion for the reduced density matrix describing
the highly-occupied polariton states ($\hbar=1$),
\begin{equation}\label{eq:1}
 \dot{\rho} = \mathcal{L}_p \rho + \mathcal{L}_d\rho -i[H,\rho],
\end{equation} where
$\mathcal{L}_p$ and $\mathcal{L}_d$ are the superoperators corresponding to pumping and 
decay respectively, while $H$ encodes the Hamiltonian dynamics of the condensates. For the 
decay we will use a sum of terms, each of the standard Lindblad form,\cite{Scully&Zubairy1997} to 
implement losses from each condensate mode. Note this assumes that each condensate mode 
emits into an independent reservoir, i.e., neglects the possibility\cite{Aleiner2012} of interference between the emission from 
different modes. 

\subsection{One reservoir pumping two modes}

\begin{figure}[ht]
\includegraphics[width=7cm]{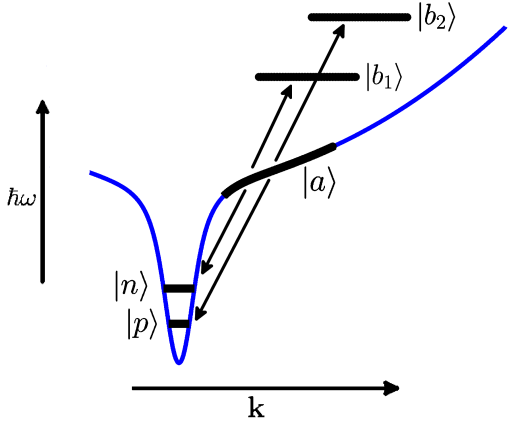}
\caption{\label{fig:1} (Color online) Schematic showing the ideal polariton dispersion
relation, and the stimulated scattering of polaritons from reservoir states, $|a\rangle$,
into low-energy condensate states, $|n\rangle, |p\rangle$, and by-product states,
$|b_1\rangle, |b_2\rangle$. The states can be localized in real space, and hence are
indicated here as involving a range of wave vectors. \label{fig:schematic}}
\end{figure}

The one reservoir, two modes problem\cite{SinghZubairy1980,Scully&Zubairy1997} is
a simplified version of the problem addressed in this paper. We use it to establish
the core formulas that we will then expand upon. We consider a reservoir of higher
energy, incoherent polaritons above the bottleneck region of the
dispersion relationship. As illustrated in Fig.\ \ref{fig:1}, we
suppose that these high energy polaritons drive condensation in two
low-energy orbitals through stimulated scattering.\cite{Deng2010}
The scattering processes will also generate particles in two,
generally different, by-product states ($|b_1\rangle,|b_2\rangle$),
which carry away the excess energy and momentum. These by-products can
be excitons, if the condensate is being
populated by exciton-exciton scattering in the reservoir, or outgoing
phonons, if it is being populated by phonon emission. Within our
approach these processes lead to the same form for
$\mathcal{L}_p$. Having two by-products allows a closed form for the dissipator
$\mathcal{L}_p$ to be obtained, by the standard procedure of
adiabatically eliminating the reservoir states.\cite{Scully&Zubairy1997}

Labeling the states of the reservoir with an index $i$, and
associating each such state with a corresponding by-product state,
gives the Hamiltonian
\begin{equation} \mathcal{H}_{p} = \sum_i g_1 a_1^\dagger
  \big(c_{b_1}^\dagger \frac{(c_a)^2}{\sqrt{2}} \big)_i + g_2 a_2^\dagger \big(c_
      {b_2}^\dagger \frac{(c_a)^2}{\sqrt{2}} \big)_i + h.c., \label{eq:scatham}
\end{equation} where $(c_a)_i$  annihilates a reservoir exciton, $(c_{b_1})_i$
and $(c_{b_2})_i$ annihilate the by-products, and $a_1$ and $a_2$ annihilate
polaritons in the condensate orbitals. $g_1$ and $g_2$ are the matrix elements
for scattering into the two condensate orbitals, which at this stage we take to
be independent of $i$. For phonon emission $(c_a)^2/\sqrt{2}$ should be replaced with $c_a$,
but reservoir levels will be traced over in the final results, and the form of
the theory is unaffected.

An important feature of the polariton dispersion is that the effective
mass, $(d^2 E/dk^2)^{-1}$, of the reservoir polaritons or excitons is several
orders of magnitude larger than that of the condensate
polaritons. Thus, the reservoir excitons are effectively immobile on
the long length scales relevant to the condensate. This is consistent
with experiments, in which the energy shifts of the polariton states,
due to the repulsion with the reservoir excitons, appear in the region
that is directly pumped.\cite{TosiBaumberg2012NatPhys, Wertz2010,
  Ferrier2011} As in the mean-field
theories,\cite{WoutersCarusoto2007,Keeling2008,Eastham2008} we may
neglect any motion of the reservoir excitons, and obtain a theory with
local gain. We take the reservoir states to be localized
orbitals at the position $r_i$, and note that the interactions
responsible for the scattering have a short range (e.g., the Bohr
radius for the exciton-exciton interaction). The scattering matrix element $g_1$, for example, is then \begin{align}
  g_1&=\int dr dr' V(r-r') \phi_1^\ast(r) \phi_{b1,i}^\ast(r')
  \phi_{a,i}(r) \phi_{a,i}(r') \nonumber \\ &\approx \int dr V_0
  \phi_1^\ast(r) \phi_{b1,i}^\ast(r) \phi_{a,i}(r) \phi_{a,i}(r)
  \nonumber \\ &\propto
  \phi_1^\ast(r_i). \label{eq:matelems}\end{align}
The spatial structure of the condensate and reservoir appears
through these matrix elements, which are proportional to the amplitude
of the condensate orbitals at the position of the reservoir.

The level structure involved in Fig.~\ref{fig:schematic} and
Eq. (\ref{eq:scatham}) is that of a two-mode laser with a
common level, $|a\rangle$, shared by the two modes. We outline
the derivation of the dissipator $\mathcal{L}_p$  for this level scheme here; similar
treatments can be found in Refs.~\onlinecite{SinghZubairy1980,Scully&Zubairy1997}.

We begin by introducing the reduced density operator describing the
condensates and one set of high-energy levels, $\rho_i =
\mathrm{Tr}_{j\neq i} \, \rho$, and its matrix elements in Fock states
\begin{equation}\label{eq:2}
\rho_i  = \sum \rho_{nmpq}^{\alpha_i \beta_i} \mid n \rangle \langle m \mid \otimes \mid p \rangle 
\langle q \mid \otimes \mid \alpha_i \rangle  \langle \beta_i \mid,
\end{equation} where $n,m$ ($p,q$) denote the occupations of the first (second) condensate mode,
and $\alpha_i,\beta_i \in \{a,b_1,b_2\}$ the occupations of the reservoir and by-product levels.
These states, for exciton-exciton scattering, have either two excitons in the reservoir
orbital $i$ (the state indicated by $a$) or one in a by-product orbital (the states indicated
by $b_1$ and $b_2$). For exciton-phonon processes, they have either one exciton in the reservoir
orbital (indicated by $a$) or a phonon in a by-product state (indicated by $b_1$ and $b_2$).
We also introduce a composite object which is the sum over these reduced density operator matrix
elements, $\rho_{nmpq}^{\alpha \beta} = \sum_i \rho_{nmpq}^{\alpha_i  \beta_i}$. The reduced Hamiltonian which carries the evolution of this density matrix is
\begin{equation}\label{eq:REDUCEDscatham}
 H_p = \sum_{i=1,2}g_i \big(a_i^\dagger c_{b_i}^\dagger c_a + c_a^{\dagger}c_{b_i}a_i \big),
\end{equation}
where we based ourselves on the phonon form of Eq. (\ref{eq:scatham}) for notational simplicity.
  
The manipulations\cite{Scully&Zubairy1997} we now present revolve around solving the following schematic relation, in the interaction picture,
\begin{equation}\label{eq:fullpump}
 \dot{\rho}=-i[H_p,\rho] + \lambda_a \rho_{[\phi \rightarrow a]} - \gamma_r \rho_{[a,b_i \rightarrow \phi]} + (\mathcal{L}_d \rho ).
\end{equation}
The $\lambda_a$ term represents the replenishing of the upper level
$|a\rangle$ from a vacuum state $|\phi\rangle$. This corresponds to
the relaxation of laser-generated higher energy polaritons into the
reservoir. The $\gamma_r$ term is for the decay from the
$|a\rangle,|b_i\rangle$ levels via channels other than the
condensates, e.g., spontaneous emission into outside cavity
modes. Linking the replenishing and relaxation processes to a common
vacuum level, $|\phi\rangle$, allows us to manipulate these
$\lambda_a,\ \gamma_r$ terms in rate equations, assuming that
the population of that level is time-independent. The rate equations
themselves correspond to the diagonal elements of the standard
Lindblad forms for the transitions shown schematically in
Eq. (\ref{eq:fullpump}). In this adiabatic limit, the levels are
eliminated, turning the replenishing into a term
$\lambda_a \rho_{nmpq}^{a a} \rightarrow r \rho_{nmpq}$, with the
effective pumping rate, $r$, given by
\begin{equation}
r = \cfrac{\lambda_a \gamma_r}{\lambda_a + \gamma_r}
\end{equation}  (see Appendix
\ref{sec:appxA} for derivation).

Second, we exploit the two different time scales associated with the processes
in Eq. (\ref{eq:fullpump}). We assume that the relaxation, proportional to $\gamma_r$,
is much faster than the dynamics of the modes due to the action of the pumping
Hamiltonian Eq. (\ref{eq:REDUCEDscatham}) and Lindblad decay, $\mathcal{L}_d$.
This allows us to solve for the slowly varying processes, $\dot{\rho} \simeq -i[H_p,\rho]$,
on a time scale for which they appear stationary, with respect to the $\gamma_r$ dynamics.
The matrix elements of the commutator of the composite object, $-i [H_p,\rho]$, are given by
\begin{eqnarray}\label{eq:3}
\dot{\rho}_{nmpq}^{\alpha \beta} &= -i \Big[ g_1 \delta_{\alpha b_1} \sqrt{n} \rho_{n-1mpq}^{a \beta} 
+ g_2 \delta_{\alpha b_2} \sqrt{p} \rho_{nmp-1q}^{a \beta}\nonumber\\
      &-g_1 \delta_{\beta a} \sqrt{m+1}\rho_{nm+1pq}^{\alpha b_1} -g_2 \delta_{\beta a}\sqrt{q
+1}\rho_{nmpq+1}^{\alpha b_2} \nonumber\\
      &+g_1 \delta_{\alpha a} \sqrt{n+1}\rho_{n+1mpq}^{b_1 \beta} + g_2 \delta_{\alpha a} \sqrt{p+1} 
\rho_{nmp+1q}^{b_2 \beta} \nonumber\\
      &-g_1 \delta_{\beta b_1} \sqrt{m}\rho_{nm-1pq}^{\alpha a}- g_2 \delta_{\beta b_2} \sqrt{q} \rho_
{nmpq-1}^{\alpha a}\Big]. \quad \quad
\end{eqnarray}
To obtain the desired equation of motion we take the trace of
Eq. (\ref{eq:3}), $\sum_{\alpha,\beta = a,a; b_1,b_1;
  b_2,b2}\dot{\rho}_{nmpq}^{\alpha \beta}=\dot{\rho}_{nmpq}$, giving a form depending on
the eight components of the density operator appearing on the
right-hand side. The equations-of-motion for these components are
obtained by using Eq. (\ref{eq:3}) once again. This second time,
we introduce the loss terms, $-\gamma_r\,\rho_{nmpq}^{\alpha \beta}$, and the effective
pumping, $r \rho_{nmpq}\delta_{\alpha a}\delta_{\beta a}$, in the right-hand side.
 
The full set of equations of motion for the eight matrix elements of
the traced density operator $\dot{\rho}_{nmpq}$ can thus be
constructed, and assembled into three matrix equations of the
form \begin{equation}\dot{R}=-MR + A.\end{equation} Here the $M$ are
$9\!\times\!9$ matrices, the $R$ are vectors formed from elements of
the density matrix, in which two quanta are being passed between the
reservoirs and the condensate modes in closed form, and the $A$
vectors are the driving terms (see Appendix \ref{sec:appxB} for
explicit forms).  The only non zero element, in each of the A vectors,
is $r \rho_{nmpq},\, r \rho_{n-1m-1pq},\, r \rho_{nmp-1q-1}$. Solving
for the three forms adiabatically, $\dot{R}=0 \Rightarrow R=M^{-1}A$,
and substituting the proper elements back into the traced version of
Eq. (\ref{eq:3}), gives the dissipator describing the pumping:
\begin{widetext}
\begin{eqnarray}\label{eq:4}
\mathcal{L}_p \rho_{nmpq} = -r \frac{ \big( g_1^2(n-m)+g_2^2(p-q) \big)^2 + \big(g_1^2(n+m
+2)+g_2^2(p+q+2) \big) \gamma_r^2 }
		      {\big( g_1^2(n-m)+g_2^2(p-q) \big)^2 + 2 \gamma_r^2 \big( g_1^2(n+m+2) + g_2^2(p
+q+2) \big) + \gamma_r^4}\rho_{nmpq} \nonumber\\
	      + \frac{ 2 r \gamma_r^2 g_1^2 \sqrt{nm} }
		      {\big( g_1^2(n-m)+g_2^2(p-q) \big)^2 + 2 \gamma_r^2 \big( g_1^2(n+m) + g_2^2(p+q
+2) \big) + \gamma_r^4} \rho_{n-1m-1pq} \nonumber\\
	      + \frac{ 2 r \gamma_r^2 g_2^2 \sqrt{pq} }
		      {\big( g_1^2(n-m)+g_2^2(p-q) \big)^2 + 2 \gamma_r^2 \big( g_1^2(n+m+2) + g_2^2(p
+q) \big) + \gamma_r^4} \rho_{nmp-1q-1}.
\end{eqnarray}
Note that $g_{1,2}$ are squared in our result; the scattering rates
into the condensates depend on the probability densities of the
condensate wavefunctions at the reservoir,
$|\phi_{1,2}(r_o)|^2$, via Eq. (\ref{eq:matelems}). Setting $n=m,\, p=q$ in Eq. (\ref{eq:4}) gives
the equation-of-motion for the population distribution,
\begin{eqnarray}\label{eq:5}
\dot{P}_{np} =\; &&
\gamma n_c \Bigg[ \dfrac{\alpha_1 n}{\alpha_1 n+ \alpha_2 (p+1)+n_s} P_{n-1p} 
	+ \dfrac{\alpha_2 p}{\alpha_1 (n+1)+ \alpha_2 p+n_s} P_{np-1} 
	- \dfrac{\alpha_1 (n+1)+ \alpha_2 (p+1)}{\alpha_1 (n+1)+\alpha_2 (p+1) +n_s} P_{np} 
\Bigg]
	\\
	&& \;\;+\gamma \Big[ (n+1)P_{n+1p} + (p+1)P_{np+1} - (n+p)P_{np} \Big], \nonumber
\end{eqnarray}\end{widetext} where we have added the standard Lindblad damping terms\cite{Scully&Zubairy1997} for radiation
from the condensate modes, assumed to decay at equal rates $\gamma$ to simplify the notation.
Here we have also introduced a dimensionless pumping parameter, $n_c=r/2\gamma$, a pump saturation
parameter, $n_s=\gamma_r^2/[4(g_1^2+g_2^2)]$, and the ratios $\alpha_{1(2)} = g_{1(2)}^2/(g_1^2+g_2^2)$.
This is the generalization, to the two-mode case, of the
pump parametrization used in Ref.\ \onlinecite{Whittaker2009}. Note
that the pumping in Eq. (\ref{eq:5}) is saturable: the gain is reduced
as the occupation increases, due to the occupation numbers in the
denominators. Furthermore it includes a gain competition effect, with
the growth rate of one mode reduced by the occupation of the
other. This arises from the common level, $|a\rangle$.

We can find a steady-state solution by requiring that the growth of an
occupation probability due to pumping matches its decay due to
loss. The first two terms on the first line of Eq.\,(\ref{eq:5})
correspond to transitions into the state of $n,p$ particles caused by
the pumping, while the final term on the second line corresponds to
transitions out of this state caused by the loss. Similarly, the final
term on the first line of Eq.\,(\ref{eq:5}) corresponds to transitions
out of the state $n,p$ caused by the pumping, while the first two
terms on the second line correspond to transitions into this state
caused by the loss. We can find a steady-state solution by requiring
that either one of these sets of rates balances. Such a detailed
balance condition\cite{Walls1974} ensures that the other set also
balances, and that there is no net flow of probability to higher or
lower occupation numbers.  Specifically, in Eq. (\ref{eq:5}), it
corresponds to the steady-state equation splitting in two identical
conditions
\begin{eqnarray}\label{eq:6} P_{np}=
 \dfrac{n_c}{n+p}\Bigg(&&\cfrac{\alpha_1 n P_{n-1p}}{\alpha_1
   n+\alpha_2 (p+1) +n_s} \nonumber\\ &&+ \dfrac{\alpha_2 p
   P_{np-1}}{\alpha_1 (n+1)+\alpha_2 p +n_s} \Bigg).
\end{eqnarray}

Note that in this single-reservoir model the gain competition leads to
single-mode behavior: except for the point $\alpha_1=\alpha_2$ the
population distribution obtained from Eq. (\ref{eq:6}) peaks when
either $n=0$ or $p=0$.\footnote{This can be seen by setting
  $\alpha_1=1,n_c>n_s$ and varying $\alpha_2$.  It is also pointed out
  in Ref.~\onlinecite{SinghZubairy1980}. In this context it may be
  helpful to define normalized quantities in terms of the coupling
  strength to one mode, rather than the total coupling, e.g.,
  $n_s^\prime=\gamma^2/(4g_1^2)$.}  Multimode behavior will become
possible when more than one reservoir is introduced, reducing the
impact of the gain competition.

We note that in the single-mode case, the population dynamics,
Eq. (\ref{eq:5}), and distribution, Eq. (\ref{eq:6}) correspond to a
standard laser-like saturable pumping. In particular, Eq. (\ref{eq:6})
reduces
to \begin{equation}\label{singlemodebalance}P_{n}=\frac{n_c}{n+n_s}P_{n-1},\end{equation}
which is Eq. (11.2.14) in Ref.~\onlinecite{Scully&Zubairy1997}; see
also Eq. (4) in Ref.~\onlinecite{Whittaker2009}. A similar recurrence
relation has been obtained by Laussy \emph{et al}.\cite{LaussyBigenwald2004}
for a model of a single condensate mode coupled to a bosonic
reservoir. Note that their Eq. (4) can be put in the form of
Eq. (\ref{singlemodebalance}) by expanding their denominator to first
order in the occupation number.

We can obtain an approximate solution to the recurrence relation,
Eq. (\ref{eq:6}), by dropping the $+1$s in the denominators, replacing
the occupation numbers with continuous variables, and approximating
$P_{n-1p} \rightarrow P_{(n,p)}-1 \cdot \partial_n P_{(n,p)}$ and
$P_{np-1} \rightarrow P_{(n,p)}-1 \cdot \partial_p P_{(n,p)}$. The
solution of the differential equation thereby obtained is the
multivariate Gaussian
\begin{multline}\label{eq:7}
P_{(n,p)} \propto \mathrm{exp} \Bigg( - \frac{(n+p)^2}{2 n_c}  + \frac{\alpha_1 n_c-n_s}{\alpha_1 
n_c}n \\ + \frac{\alpha_2 n_c-n_s}{\alpha_2 n_c}p \Bigg).
\end{multline} 

\subsection{Many reservoirs pumping many modes}\label{sec:many2many}

We now generalize the above results to allow many different reservoirs to pump many different
condensate modes. The Hamiltonian for scattering from each of these reservoirs is Eq.\,(\ref{eq:scatham}),
generalized to many condensate modes. We now use $g_{ij}$ to denote the matrix element for a transition
from one of the initial $(|a\rangle)_i$ levels in reservoir $i$ into a by-product state $(|b_j\rangle)_i$
in that reservoir, and a polariton in a condensate mode $j$. Note that each reservoir generally involves
many high-energy states, as in Eq.\,(\ref{eq:scatham}), whose matrix elements for scattering into a particular
condensate mode are all supposed to be approximately equal. We further assume that the different reservoirs
are independent, except for their coupling via the condensates. From the forms of the matrix
elements, Eq.\,(\ref{eq:matelems}), we see that the first assumption is valid when the reservoirs
are regions of space that are small enough for the variation in the condensate wavefunctions across 
each to be ignored. The second condition requires that the reservoirs are large compared with the mean
free path of the high-momentum excitons above the bottleneck. 

As an aper\c{c}u,\cite{RacineThesis}
the generalization to many reservoirs means that the full high-energy subspace is
subdivided into blocks. The reservoirs and by-products density matrix elements, defined after Eq. (\ref{eq:2}), now carry
indices for each reservoir, $\rho_{..}^{\alpha_1 \beta_1 \alpha_2 \beta_2.. \alpha_i \beta_i..}$, and
the matrices $M$ become Kronecker sums, $M = \sum_i I \otimes.. \otimes M_i \otimes ..\otimes I$. But the assumed independence of the different reservoirs means that the pumping terms ($A$ vectors)
appear in each subspace, and the result is that the dissipator for pumping by many reservoirs is the
sum of dissipators for individual reservoirs.  The generalization to many condensates means that
each $M_i$ associating the $i^{th}$ reservoir with $v$ condensates is an $(1\!+\!v)^2\!\times\!(1\!+\!v)^2$
matrix. The forms of these matrices, however, allow the relevant elements of\begin{widetext}their inverses to be
obtained [most of the terms in the cofactors and the determinants which form $(M_i)^{-1}_{i1}\!=\!\tfrac{(C_i)_{i1}}{|M_i|}$ cancel],
and the final result is a direct generalization of Eq. (\ref{eq:4}) : 
\begin{eqnarray}\label{eq:8}
\mathcal{L}_p \rho_{n_1\, n_2\;..}^{m_1m_2..} = \sum_i r_i \Bigg[\sum_j &\dfrac{ 2 \gamma_r^2 g_{ij}
^2 \sqrt{n_j m_j} \rho_{n_1\, n_2\;...n_j-1..}^{m_1m_2..m_j-1..} }
			{\big(\sum_k g_{ik}^2(n_k-m_k)\big)^2 + 2 \gamma_r^2 \sum_k g_{ik}^2(n_k+m_k+2) 
-4 \gamma^2_r g_{ij}^2 + \gamma_r^4} \nonumber\\
			&- \dfrac{ \big(\sum_j g_{ij}^2(n_j-m_j)\big)^2 + \gamma_r^2 \sum_j g_{ij}^2(n_j+m_j
+2)  }
			{\big(\sum_j g_{ij}^2(n_j-m_j)\big)^2 + 2 \gamma_r^2 \sum_j g_{ij}^2(n_j+m_j+2) + 
\gamma_r^4} \rho_{n_1\, n_2\;..}^{m_1m_2..} \Bigg]
\end{eqnarray}\end{widetext}
($\rho_{n_1\,n_2\;..}^{m_1m_2\;..}\equiv\langle n_1,n_2,..|\rho|m_1,m_2,..\rangle$,
with $n_j$ and $m_j$ occupation numbers of the condensate modes). Note that the growth of condensate $j$
due to reservoir $i$ depends on the (squared) magnitude of the condensate orbital at the position of that
reservoir, via the matrix element $g_{ij}$, as well as the pump rate for that reservoir, $r_i$. Since the
reservoir $i$ is now feeding many condensate modes, the gain is reduced according to all their occupations,
giving rise to the sums over condensate modes in the second terms of the denominators.

The generalization of the dimensionless pumping parameter for reservoir $i$, introduced above,
is $n_i^c = r_i/2\gamma$, while the pump saturation parameter becomes $n_i^s = \gamma_r^2/(4 \sum_j g_{ij}^2)$,
and the normalized transition strengths  become
\begin{equation}\alpha_{ij} = g_{ij}^2/\sum_k g_{ik}^2.\label{eq:ntstrengths}\vspace{0.5in}\end{equation}
\vspace{-1cm}

Finally, the population dynamics equation (\ref{eq:5}) now reads
\begin{eqnarray}\label{diffP}
\dot{P}_{n_1 n_2..}&&= \gamma \sum_i n_i^c \sum_j \Big[ \frac{\alpha_{ij} n_j P_{n_1n_2..n_{j}-1..}}
{\sum_k \alpha_{ik}(n_k+1) -\alpha_{ij}+ n_i^s } \nonumber\\
 &&- \frac{\alpha_{ij} (n_j+1) P_{n_1 n_2..}}{\sum_k \alpha_{ik}(n_k+1)+n_i^s} \Big] \\ 
 &&+ \gamma \sum_j \Big[ (n_j+1)P_{n_1 n_2.. n_j+1..} - n_j P_{n_1 n_2..} \Big].\nonumber
\end{eqnarray} We have found an approximate steady-state solution of this equation, for the special case where
the coupling ratios among the $u$ reservoirs and $v$ condensates obey
$\alpha_{ii+j} = \beta_j, i=1,2..u, j=0,1,2,..,v-1$, with the index $i+j$ treated circularly
around $v$, and where all the $n_i^s, n_i^c$ are the same. The solution, valid to the extent that Eq. (\ref{eq:error}) holds,
is the multivariate Gaussian
\begin{eqnarray} \label{eq:10}
P&&_{(n_1,n_2,..)} \propto 
	 \mathrm{exp} \Bigg[\!\! -\frac{1}{2} \begin{pmatrix} n_1 & n_2 &.. \end{pmatrix} \cdot 
\mathbb{S} \cdot
		\begin{pmatrix} n_1 \\ n_2 \\ .. \end{pmatrix} 
		+ \mathbb{Z} \cdot \begin{pmatrix} n_1 \\ n_2 \\ ..\end{pmatrix} \!\Bigg] \nonumber \\ 
\end{eqnarray}
with
\begin{equation}\label{analA}
\mathbb{S}=\begin{pmatrix} \frac{\sum_i \beta_i^2}{n_c} & \frac{\sum_{i,j\neq i} \beta_i \beta_{j} }{n_c} & ..\\ \frac{\sum_{i,j\neq i} \beta_i \beta_{j}}{n_c} & \frac{\sum_i \beta_i^2}{n_c} & .. \\
 ..&..&..  \end{pmatrix} 
\end{equation}
and
\begin{equation}\label{analB}
\mathbb{Z} = \begin{pmatrix} \tfrac{n_c-n_s}{n_c} & \tfrac{n_c-n_s}{n_c} &.. \end{pmatrix}.
\end{equation} The validity of Eq.~(\ref{eq:10}) can be seen by substituting it into the continuous version of Eq.~(\ref{diffP}). For the case of two condensates and two reservoirs this gives
\begin{eqnarray}\label{eq:error}
0 \approx  \beta_o \beta_1\! \Bigg( \cfrac{ (\beta_1\!-\!\beta_o)(n\!-\!p)}{\beta_o n\! +\!\beta_1 p\!+\!n_s} +
 \cfrac{ (\beta_o\!-\!\beta_1)(n\!-\!p)}{\beta_1 n\!+\!\beta_o p\! +\!n_s} \Bigg).
\end{eqnarray}

\section{\label{sec:coherencefun} Coherence functions}

\subsection{Low-energy Hamiltonian}
\label{sec:lowenh}

We now consider how the dissipative dynamical model, described above,
may be further developed, and used to calculate the dynamical
characteristics of a multimode polariton condensate. In particular, we
consider the calculation of time-dependent first and second order
coherence functions. To keep the notation manageable we consider
explicitly a two-mode condensate; the formulation is such that the
generalization is reasonably straightforward.

The first step is to introduce the Hamiltonian dynamics of the
condensate modes. In particular, we include the polariton-polariton interactions.
The underlying interaction~\cite{Ciuti2000} is predominantly due to the exchange
terms in the Coulomb and radiative couplings (phase-space filling), so that its
range is on the order of the Bohr radius. It should thus, in the present low-energy
theory, be understood to be a contact interaction, $V(r)=V_0\delta^d(r)$, with matrix element
$\mu_{ijkl}=V_0 \int d^d r \psi^\ast_i(r)\psi_j(r)\psi^\ast_k(r)\psi_l(r) $.
The low-energy Hamiltonian is then
\begin{equation} H=\sum_{i=1,2} \omega_i a^\dagger_i a_i +
  \tfrac{1}{2} \sum_{ijkl}
  \mu_{ijkl}a^\dagger_ia_ja^\dagger_ka_l, \label{eq:genham}\end{equation}
where $\omega_i$ are the single-particle energies. This form is
obtained by diagonalizing the single-particle Hamiltonian, so that
$a_i$ is the annihilation operator in a single orbital. An example of
this procedure, diagonalizing the Hamiltonian for a double-well system
by eliminating the hopping term, is described in
Sec.~\ref{sec:twomodes}.

In order to render the general interacting bosonic Hamiltonian,
Eq. (\ref{eq:genham}), tractable, we consider the case in which there
is some trapping potential creating localized single-particle orbitals
for the polaritons. We furthermore assume the strong-trapping limit,
where the energy separation between the orbitals, $\omega_i-\omega_j$,
is large compared with the interaction energy (of order $V_0n$ for an
overall condensate density $n$). We will therefore neglect the parts
of the interaction which transfer particles between different
orbitals, i.e., non resonant scattering processes between condensates
in different modes. This makes the problem tractable since the
equation of motion becomes diagonal in Fock space. It is consistent
with the semiclassical limit of the equations of motion, in which such
processes give rise to rapidly oscillating terms.\cite{Eastham2008}
However, the nonlinear terms
which conserve the number of particles in each orbital must be
retained, since their effects are not suppressed by the differences in
single-particle energies. For the two-mode problem we will thus take
for the interaction Hamiltonian the form
\begin{equation}\label{HNL}
H_{(NL)} = \kappa (a_1^\dagger a_1)^2 + \kappa (a_2^\dagger a_2)^2 + \eta a_1^\dagger a_1 
a_2^\dagger a_2,
\end{equation} including the Fock space diagonal interactions within each mode
(strength $\kappa$) and between the modes (strength $\eta$). However, the Bogoliubov
or parametric scattering terms, such as $a_1^\dagger a^\dagger_1 a_2 a_2$, are neglected.
Since the single-particle energies are just energy shifts in the following they will often
be dropped.  The form of the equation of motion, (\ref{eq:1}), we consider is thus
\begin{eqnarray}
\dot{\rho} = \mathcal{L}_p \rho&&\ -\ \tfrac{\gamma}{2}\bigl( (a_1^{\dagger}a_1 + a_2^{\dagger}
a_2)\rho + \rho(a_1^{\dagger}a_1 + a_2^{\dagger}a_2) \nonumber\\
&&\ -\ 2a_1\rho a_1^{\dagger} - 2a_2\rho a_2^{\dagger} \bigr) - i[H_{(NL)}, \rho].\label{eq:gentwomodes}
\end{eqnarray}

We note that the form of the interactions in Eq. (\ref{HNL}) do not enter into the equation for the population dynamics, Eqs. (\ref{eq:5}, \ref{diffP}), since $H_{(NL)}$ commutes with the diagonal elements of $\rho$. Their role will be primary however in the calculation of coherence functions below.

\subsection{First-order coherence functions}

A key application of the equation of motion, (\ref{eq:1}), is to
calculate the linewidths of the emission from the individual
condensate modes.~\cite{Whittaker2009} Equivalently, one can consider
the first-order coherence function of the emission from mode 1, for
example, $g^{(1)}(\tau)\propto \langle a_1^\dagger(\tau)
a_1(0)\rangle$. By the Wiener-Khinchin theorem, the Fourier transform
of this correlation function gives the power spectrum of the
electromagnetic field, i.e., the emission
spectrum.\cite{loudonspectrum} Such spectra have been studied
experimentally.\cite{Love2008} The linewidth has been shown to be
generated by the interplay between interactions and the polariton
number fluctuations, discussed above, which together imply energy
fluctuations.

The first-order coherence function may be calculated from the
equation of motion for the reduced density matrix, by exploiting a
form of the quantum regression
theorem.\cite{Scully&Zubairy1997,Swain1981} Thus, the two-time
correlation function $\langle a_1^\dagger(\tau)a_1(0)\rangle$ is the
expectation value $\langle a_1^\dagger \rangle=\Tr a_1^\dagger
\rho^\prime$, with a density operator
$\rho^\prime(\tau)=e^{\mathcal{L}\tau}\rho^\prime(0)$ obtained by evolving
$\rho^\prime(0)=a_1\rho(0)$ according to Eq. (\ref{eq:1}). In this form
the regression theorem holds provided the system-reservoir coupling is
weak, so that the full density operator factorizes. This is already
implicit in our model for pumping and decay. We note that there is
also a stronger version of the quantum regression theorem, which
relates the equations of motion for $n$-time correlation functions to
those for $m$-time ones.\cite{Scully&Zubairy1997,Ford1996} This is not
useful here, because the nonlinearities mean that $\langle
a_1(t)\rangle$, for example, does not obey a closed set of linear
equations of motion.

To implement this quantum regression approach, we introduce the
distribution\cite{Whittaker2009} $u_{np}(\tau)=
\sqrt{n}\rho^\prime_{n-1npp}(\tau)$
, so that \begin{equation}g^{(1)}(\tau)\propto \langle
  a_1^\dagger \rangle=\sum_{n,p}
  u_{np}(\tau).\label{eq:g1qrt}\end{equation} The initial condition
for the evolution is
\begin{equation}\label{eq:uinicond}
  u_{np}(0)=\langle n,p|a^\dagger_1 a_1\rho(0)|n,p\rangle=nP^{ss}_{np},\end{equation} where the steady-state population
distribution $P^{ss}_{np}$ is obtained from Eq. (\ref{eq:10}) [or more generally Eq.~\ref{diffP}]. From Eq. (\ref{eq:1}) we find the equation of motion for $u_{np}$,
\begin{eqnarray}\label{linewidtheq}
\dot{u}_{np} = \gamma \sum_{i=1,2} n^c_i &&\bigg[ \dfrac{\alpha_{i1}n}{\alpha_{i1}(n-\frac{1}{2}) + 
\alpha_{i2}(p+1) + n^s_i} u_{n-1p} \nonumber\\
		&&+ \dfrac{\alpha_{i2} p}{\alpha_{i1}(n+\frac{1}{2}) + \alpha_{i2} p + n^s_i} u_{np-1} 
\nonumber\\
		&&- \dfrac{\alpha_{i1}(n+\frac{1}{2}) + \alpha_{i2}(p+1)}{\alpha_{i1}(n+\frac{1}{2}) + 
\alpha_{i2}(p+1) +n^s_i} u_{np} \bigg] \nonumber\\
		+\gamma \bigg[ n u_{n+1p} &&+ (p+1) u_{np+1} - \big( (n -\frac{1}{2}) + p \big) u_{np} 
\bigg] \nonumber\\
		+i \big[ \kappa (2n-&&1) + \eta p \big] u_{np}.
              \end{eqnarray}
We have neglected terms $\mathcal{O} (\frac{n}{n_s} \cdot n)$, since $n_s \gg \langle n \rangle$ for our system.
The solution to Eq. (\ref{linewidtheq}),
with the appropriate initial condition, Eq. (\ref{eq:uinicond}),
gives the first-order coherence function via Eq. (\ref{eq:g1qrt}). For 
up to several hundreds of particles, $\dot{u}_{np}$ can be integrated
numerically to obtain $g^{(1)}(\tau)$.

In the single-mode case, an expansion based on the one mode equivalent of Eq. (\ref{linewidtheq})
was solved analytically in Ref.\ \onlinecite{Whittaker2009}. A Kubo form\cite{Hamm2005,Kubo1954}
was reached which incorporates the interaction and the slower Schawlow-Townes broadening. Within each
term both a Lorentzian and a Gaussian lineshape can be obtained depending on which limit is applied,
i.e., motional broadening and the static limit. For multimode condensation, we could not reach a simple
analytic formula for the first-order correlation function and we therefore resort to the semianalytic
approaches presented below.

\subsection{First-order coherence: Fokker-Planck approach}

To make Eq. (\ref{linewidtheq}) tractable more widely, we recast it in
the form of a soluble partial differential equation. This process follows
the derivation of the Fokker-Planck equation for a one-step
Markov process.\cite{VanKampen} It involves approximating the occupation numbers by
continuous variables, and expanding the finite-difference operators in
terms of differentials (Kramers-Moyal expansion). 

We first introduce step operators $\mathbb{E}_{n,p}$, whose action,
for example, is $\mathbb{E}_{n}nu_{np}=(n+1)u_{n+1 p}$. This allows us
to rewrite Eq. (\ref{linewidtheq}) as
\begin{eqnarray}
 \dot{u}_{np} =&& (\mathbb{E}_n^{-1}-1,\mathbb{E}_p^{-1}-1)\cdot\big[(q_n,q_p)u_{np}\big] 
\nonumber\\
		&&+\ (\mathbb{E}_n-1,\mathbb{E}_p-1)\cdot\big[(r_n,r_p)u_{np}\big] \nonumber\\
		&&+\ (h+s)u_{np}\label{eq:meqform}
\end{eqnarray}
where
\begin{eqnarray}\label{prefac1}
&&q_{n,p}=\ \gamma n_c \Big( \frac{\beta_{o,1}}{D_o} + \frac{\beta_{1,o}}{D_1} \Big)(n,p+1) \\
&&r_n=\ \gamma (n-1),\quad r_p=\ \gamma\, p.
\end{eqnarray} The first two lines in Eq.~(\ref{eq:meqform}) correspond to the master equation for
a one-step stochastic process, such as nearest-neighbor transitions in two
dimensions, with $u_{np}$ interpreted as a probability, and
$q_{n,p},r_{n,p}$ transition rates. These contributions conserve the sum
of $u$, and alone would lead to the Fokker-Planck (i.e., continuity)
equation in the continuous limit. A nonconserving term proportional
to
\begin{equation}h = \frac{\gamma n_c}{2} \Big( \frac{\beta_o}{D_o}
  + \frac{\beta_1}{D_1} \Big)- \frac{\gamma}
  {2}, \label{eq:hprefac}\end{equation}
remains when the gain and loss terms in Eq.~(\ref{linewidtheq}) are recast in this way. Here $\beta_o
= \alpha_{11},\alpha_{22}$ and $\beta_1 = \alpha_{12},\alpha_{21}$,
and
\begin{equation}
D_{o,1} = \beta_{o,1}(n+\tfrac{1}{2}) + \beta_{1,o}(p+1) +n_s.
\end{equation}
The interactions, also, lead to a non-conserving term, proportional to  
\begin{equation}
s = i\big[ \kappa (2n-1) + \eta p\big]. \label{sprefac}
\end{equation}
We then Taylor-expand the step operators
$\mathbb{E}_{n,p}^{\pm1}\rightarrow
1\pm\partial_{n,p}+\frac{1}{2}\partial_{n,p}^2$, and obtain the
continuous approximation
\begin{eqnarray}
\dot u_{(n,p)} = &&(h+s) u_{(n,p)}+ \nabla \cdot [(\vec{r}-\vec{q})u_{(n,p)}] \nonumber \\ && + \tfrac{1}{2} \nabla^2 \cdot [(\vec{r}+\vec{q})u_{(n,p)}] \nonumber \\ 
= &&(h+s) u_{(n,p)} \nonumber \\ &&+ \nabla \cdot \{[\vec{r}-\vec{q}+\tfrac{1}{2} \nabla\odot (\vec{r}+\vec{q})]u_{(n,p)} \nonumber \\ && \qquad + \tfrac{1}{2} (\vec{r}+\vec{q})\odot \nabla u_{(n,p)}\},\label{cont_u}
\end{eqnarray} where
$\vec{r}=(r_n,r_p), \vec{q}=(q_n,q_p), \nabla=(\partial_n,\partial_p)$, $\nabla^2=(\partial^2_n,\partial^2_p)$,
and $\odot$ denotes elementwise multiplication. We notice that the conserving part of Eq. (\ref{cont_u})
is a convection equation in the occupation number space, with a position-dependent drift velocity, given by
the prefactor of $u$ inside the divergence, and a diffusion coefficient, given by the prefactor of $\nabla u$.
The remaining terms, proportional to $h+s$, cause the integral of $\dot{u}$ to be non-zero. They induce the
decay in magnitude of $u$, and hence are responsible for
decoherence and the finite linewidth, according to Eq. (\ref{eq:g1qrt}). The standard Schawlow-Townes
linewidth arises from the term proportional to $h$, while the
interaction broadening arises from that proportional to $s$.

Equation (\ref{cont_u}), like the Fokker-Planck equation, is soluble when
$h$, $s$, and the drift coefficients are at most linear functions of
the occupation numbers, and the diffusion coefficients are
constants. We therefore expand these coefficients appropriately in
Taylor series around the mean of the initial conditions,
$n_l=\frac{\langle nu_o\rangle}{\langle u_o\rangle}$,
$p_l=\frac{\langle pu_o\rangle}{\langle u_o\rangle}$. For simplicity
we also neglect the quadratic terms in the expansion of
$\vec{r}+\vec{q}$, which would contribute a small linear term to the
drift coefficient.

We define $n^\prime=n-n_l$, and introduce the
reciprocal representation $g=\iint
e^{-i(k_1,k_2)\cdot(n'\!,p')}u_{(n'\!,p')}\mathrm{d}n'\mathrm{d}p'$. The
appropriately linearized form of Eq. (\ref{cont_u}) is then
\begin{eqnarray}\label{LaplaceEq}
 \dot{g} + \sum_j \big( a_j + \sum_i && b_{ij} k_i \big) \partial_{k_j}g
 = \nonumber
 \\ && cg + \sum_i d_i k_i g + \sum_{i} e_{i} k_i^2 g,
\end{eqnarray} with coefficients
\begin{eqnarray}
A=&& -i\nabla(h+s)|_{n_l,p_l} \label{coefA}\\
B=&& \left.\begin{pmatrix} \nabla(r_n-q_n) \\ \nabla(r_p-q_p) \end{pmatrix} \right|_{n_l,p_l}\\
c=&& h|_{n_l,p_l}\\
D=&& i\big(r_n-q_n+\tfrac{1}{2}\partial_n(r_n+q_n), \nonumber\\&& \quad\quad r_p-q_p+\tfrac{1}
{2}\partial_p(r_p+q_p)\big) \big|_{nl,pl} \label{Dterm}\\
E=&&-\frac{1}{2}\left.\begin{pmatrix}r_n+q_n&0\\0&r_p+q_p\end{pmatrix}
\right|_{n_l,p_l} \hspace{-0.5cm}, \label{coefE}
\end{eqnarray} where $D=(d_1,d_2)$, \emph{etc.}. The origins of the
various terms in Eqs. (\ref{coefA})--(\ref{Dterm}) can be seen in
Eq. (\ref{cont_u}); note there are contributions from both $r-q$ and
$r+q$ to the lowest-order drift matrix $D$, from the two
terms on the second-to-last line of Eq. (\ref{cont_u}). At
$k_{1,2}=0$, the solution of Eq. (\ref{LaplaceEq}) shall give $g^{(1)}(\tau)$.

We solve Eq. (\ref{LaplaceEq}) using the method of characteristics\
\cite{VanKampen} to reduce it to a set of coupled ordinary
differential equations. The characteristic equation is $\dot{K} = A +
K B$, which may be solved in the eigenbasis of $B$, $\mathbb{B} =
P^{\,\sminus1}BP$, where $P$ is the eigenvector matrix of $B$, before
transforming back to the original basis. This gives
\begin{eqnarray}
K_o = -AB^{\sminus1}+(AB^{\sminus1} + K) e^{\sminus B\tau},
\end{eqnarray} where $K_0=K(\tau=0)$.
Note that the origin $\tau=0$ is significant and the constant
$-AB^{\sminus1}$ cannot be dropped. We then obtain
\begin{equation}
g = g_o(K_o)\, \mathrm{exp}\Big[c \tau + D\!\int_0^\tau K \mathrm{d}t' + \int_0^\tau K E \tilde{K} \mathrm{d}t' 
\Big],
\end{equation} using $\tilde{K}$ to denote the transpose of $K$.
The first integral inside the exponential is 
\begin{eqnarray}
\int_0^\tau K \mathrm{d}t'\ &&= -A B^{\sminus1}\tau + (K_o+AB^{\sminus1}) B^{\sminus1}(e^{B\tau}\!-\!I)\nonumber\\
		      =-&& A B^{\sminus1}\tau + (AB^{\sminus1}+K)B^{\sminus1}(I\!-\!e^{\sminus B\tau}),
\end{eqnarray}
which at $K=0$ gives
\begin{equation}
 \zeta(\tau) = -DAB^{\sminus 1}\tau + DAB^{\sminus 1}B^{\sminus1}(I\!-\!e^{\sminus B\tau}).
\end{equation}
  The second integral is solved in a similar fashion, only in this
  case the result depends explicitly on the eigenvalues and
  eigenvectors of $B$. We define $E_{p} = P^{\,\sminus1}E \tilde{P}^{\,\sminus1}$ and obtain
\begin{eqnarray}\label{secondInt}
&&\xi(\tau)= AB^{\sminus1}E \tilde{B}^{\sminus1}\tilde{A}\tau \nonumber\\
&&\quad- \Big( AB^{\sminus1}B^{\sminus1}(I\!-\!e^{\sminus B\tau})E \tilde{B}^{\sminus1}\tilde{A} + \mathrm{T.c.} \Big)
+ AB^{\sminus1}P  \nonumber\\ &&\quad \times\Big[ 
      \sum_{ij} \hat{\imath} \otimes \hat{\jmath} \frac{E_{p\,ij}}{\mathbb{B}_{ii}\!+\!\mathbb{B}_{jj}}(I\!-\!e^{-(\mathbb{B}_{ii}+\mathbb{B}_{jj})\tau})
      \Big] \tilde{P} \tilde{B}^{\sminus1} \tilde{A}.
\end{eqnarray}
The tensor product $\hat{\imath} \otimes \hat{\jmath}$ with unit vectors $\hat{\imath},\hat{\jmath}$ generates a matrix with all zeros except at position $ij$.
We finally obtain the solution 
\begin{equation}
 g^{(1)}(\tau)= g_o(\tau)\,\mathrm{exp}\big(c \tau + \zeta(\tau) +
 \xi(\tau)\big). \label{eq:g1result}
\end{equation}
The function $g_o(\tau)$ is related to the transform of $u_{np}(0)$,
given by Eq. (\ref{eq:uinicond}),
\begin{equation}\label{initalFP}
g_o(\tau)  = \Big(\int e^{-iK'_o\cdot (n,p)} n P_{(n,p)} \mathrm{d}n\,\mathrm{d}p\Big)
	\mathrm{exp}(i K'_o \cdot (n_l,p_l)),
\end{equation}
where $K'_o = -AB^{\sminus1}(I-e^{\sminus B\tau})$ is $K_o$ at
$K(\tau)=0$. We also shifted the transform by the linearization points,
$n'$\,$=$\,$n$\,$-$\,$n_l$, $p'$\,$=$\,$p$\,$-$\,$p_l$, such that it
is consistent with the rest of the solution. The coefficients, Eqs.
(\ref{coefA}$\,$-$\,$\ref{coefE}), can be obtained analytically, while
the diagonalization of $B$ and integral in $g_0$ are performed
numerically.

\subsection{First-order coherence: static limit}

\label{sec:first-order-coher}

For sufficiently short timescales the gain and loss processes will not
change the occupation numbers of the condensate orbitals. We can
therefore calculate $g^{(1)}(\tau)$ allowing only for the effects of
the nonlinear Hamiltonian, Eq. (\ref{HNL}). This is the static-limit
calculation previously discussed for the single-mode
case.\cite{Whittaker2009} The time scale over which it is valid is the
time scale for intensity fluctuations, given by the decay of
$g^{(2)}(t)$; due to critical slowing down this timescale becomes long
close to threshold, and most of the decay of $g^{(1)}(\tau)$ is
captured correctly. For the two-mode case we find
\begin{eqnarray}
|g_s^{(1)}(\tau)| &&= \Big|\Tr \sum_{np} a_1^\dagger e^{-iH_{(NL)}\tau}\Big(P_{np}\,a_1|n,p\rangle 
\langle n,p|\Big) \nonumber \\ && \qquad\quad\qquad\qquad\qquad\qquad\qquad\qquad \times e^{iH_{(NL)}\tau} \Big| \nonumber\\
	    && \propto \Big|\iint \! P_{(n,p)}e^{i(2\kappa n + \eta p)\tau }\,\mathrm{d}n\,\mathrm{d}p \;\Big|,
          \end{eqnarray} where in the second line we have approximated a factor of $n$ inside the
integral as a constant and treated the occupation numbers as continuous. Due to the cutoff in  $P_{(n,p)}$
the integral will be performed numerically. Notice that the kernel in the integral indeed corresponds
to the interaction term proportional to $s$ in Eq. (\ref{linewidtheq}); all the other dynamics is neglected.

\subsection{Coherence of Josephson oscillations}

The approaches described above for the first-order coherence function
can be generalized to calculate higher-order correlation functions. An
interesting example is the second-order cross-correlation function
$g^J(\tau) \propto \langle a_1^\dagger(\tau) a_2(\tau) a_2^\dagger a_1
\rangle$. This function characterizes the coherence of the intensity
oscillations, caused by the beating between the different emission
frequencies in a multimode condensate. They can be interpreted as a
form of Josephson oscillation.~\cite{Leggett2001}

To motivate the consideration of the correlation function $g^J(\tau)$
we consider the mean-field limit where the creation and annihilation
operators can be treated as classical oscillating variables,
$a_{1,2}(t)\to\sqrt{n_{1,2}} e^{i(\omega_{1,2} t+\phi_{1,2})}$, in
other words the eigenvalues of coherent states,
$|\alpha_{1,2}\rangle$.~\cite{Leggett2001, Scully&Zubairy1997} For two
orbitals with amplitudes $c_1,c_2$ at a point $r$, the density or
intensity is $n=(c_1 a_1^\dagger+c_2
a_2^\dagger)(c_1a_1+c_2a_2)$. Thus, in that theory, there is an
oscillating contribution to the intensity, proportional to \begin{equation}\cos
[(\omega_1-\omega_2 )t+\phi_1-\phi_2].\label{eq:densosc}\end{equation}

The intensity oscillations are not zero in the mean-field theory
because it assumes a well-defined phase for the condensates, and hence
a well-defined phase difference between them. However, in the
strong-trapping limit considered here there are no terms in the
Hamiltonian which fix this relative phase, and the averaged intensity
does not oscillate, $\langle a_1^\dagger(\tau) a_2(0) \rangle = 0
$. Even in the absence of phase-fixing terms in the Hamiltonian,
however, a phase would arise in each member of an ensemble, i.e., a
single run of an experiment, due to spontaneous symmetry breaking. The
fluctuations of the phases between members of the ensemble account for
the vanishing of the oscillations on average. We can nonetheless study
how the oscillations in each member behave, by considering the
correlation function of the intensity, \begin{align}\langle
  n(\tau)n(0)\rangle=
  \langle&(a_1^\dagger(\tau)+a_2^\dagger(\tau))(a_1(\tau)+a_2(\tau))\nonumber\\&\;\times(a_1^\dagger
  + a_2^\dagger)(a_1+a_2)\rangle,\end{align} and in particular the
component at the beat frequency $\langle a_1^\dagger(\tau) a_2(\tau)
a_2^\dagger a_1\rangle$. (We omit the amplitudes $c_{1,2}$ above for
notational simplicity.)

The calculation of $g^{J}(\tau)$ closely parallels that of the first-order
coherence. We again use the quantum regression theorem, so that
$g^{J}(\tau)$ is the average $\langle a_1^\dagger a_2 \rangle$, with
the density operator obtained by evolving $a_2^\dagger a_1 \rho(0)$
over the time $\tau$. We introduce the distribution
\begin{equation}
u_{np}^J = \sqrt{n}\sqrt{p+1} \, \rho'_{n-1np+1p}(\tau),
\end{equation} such that $g^{J}(\tau)=\sum_{n,p}u_{np}^J$. The initial
condition is $u_{np}^J(0) = n(p+1) \,
P_{np}^{ss}$, and the evolution obeys
\begin{eqnarray} \label{eq:josephsonueq}
\dot{u}^J_{np} = \gamma \sum_i &&n^c_i \bigg[ \frac{\alpha_{i1}n}{\alpha_{i1}(n-\frac{1}{2}) + 
\alpha_{i2}(p+\frac{3}{2}) + n^s_i} u^J_{n-1p} \nonumber\\
			&&\quad + \frac{\alpha_{i2} (p+1)}{\alpha_{i1}(n+\frac{1}{2}) + \alpha_{i2}(p+\frac{1}{2}) 
+ n^s_i} u^J_{np-1} \nonumber\\
			&&\quad - \frac{\alpha_{i1}(n+\frac{1}{2}) + \alpha_{i2}(p+\frac{3}{2})}{\alpha_{i1}(n+\frac{1}
{2}) + \alpha_{i2}(p+\frac{3}{2}) +n^s_i} u^J_{np} \bigg]\nonumber\\
		+\gamma \bigg[ n &&u^J_{n+1p} + (p+1) u^J_{np+1} - \big( n + p \big) u^J_{np} \bigg] \\
		+\: i \Big( \kappa &&(2n-1) - \kappa (2p+1) + \eta (-n+p+1) \Big) u^J_{np}. \nonumber
\end{eqnarray}
The static limit can be calculated in the same fashion as above and
gives the expression
\begin{equation} \label{eq:josephsonstatic}
|g_s^J(\tau)| \propto \Big|\iint \! P_{(n,p)}e^{i(2\kappa - \eta)(n - p)\tau }\,\mathrm{d}n\,\mathrm{d}p\Big|.
\end{equation}

\section{\label{sec:twomodes} Condensation in a double-well potential}

\begin{figure}[t!]
\includegraphics[width=8.5cm]{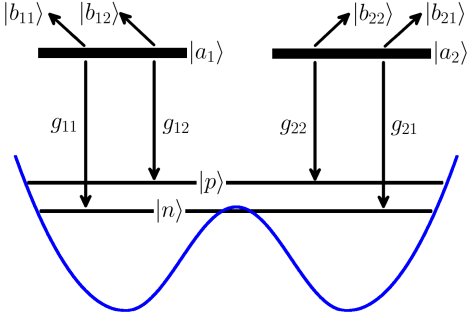}
\caption{\label{fig:2} (Color online) Model for condensation in a double-well potential,
in which two orbitals, extended across the well due to tunneling, are pumped by independent
reservoirs localized on the left and right. The occupations of the two delocalized orbitals
are denoted $n$ and $p$.\label{fig:twowellcartoon}}
\end{figure}

We now apply the general theory developed above to the specific
problem of polariton condensation in a double-well potential, with
incoherent pumping provided by a high-energy reservoir. This system,
illustrated in Fig.~\ref{fig:2}, is a form of Josephson
junction. Oscillations of the density of polaritons in each well,
analogous to the a.c. Josephson effect, have recently been observed
experimentally.\cite{Lagoudakis2010} They are due to the presence of
two highly occupied states of different energies, as predicted by
Wouters \cite{Wouters2008} and by Eastham.\cite{Eastham2008} The
theory developed above will allow us to obtain the quantum statistical
properties of the light emitted from this type of double-well system,
including the linewidths and the dephasing time of the density
oscillations.

We note that, in addition to their observation in a system with
continuous incoherent pumping,\cite{Lagoudakis2010} where losses are
compensated by gain, Josephson oscillations have also been observed in
transient condensates.\cite{BlochJoseph2013,Raftery2014} Such
condensates are created by an initial excitation pulse, and the
Josephson effects are observed following the pulse, but before the
condensate decays. Abbarchi \emph{et al.}\cite{BlochJoseph2013} studied the
dynamics of polaritons created by ultrafast resonant excitation in a
double-well potential, and observed linear and nonlinear oscillations
as well as macroscopic quantum self-trapping (MQST). The condensates
in this case are generated directly by a pump laser, rather than by
scattering from an incoherent reservoir. Raftery \emph{et al}.\cite{Raftery2014} studied the dynamics of interacting photons in
coupled superconducting resonators; they observed linear and nonlinear
oscillations, collapses and revivals reflecting quantum effects, and a
dynamical transition to MQST as the population decays. Josephson
effects for polaritons with continuous \emph{coherent} pumping have
also been considered theoretically,\cite{sarchi2008} as have some
Josephson phenomena for polaritons with incoherent gain and
loss.\cite{MagnussonShelykh2010,read2010}

As shown in Fig.~\ref{fig:2}, we consider a situation in which the two
lowest orbitals of a double-well potential are being pumped by two
reservoirs, one for each of the wells. We begin with a tight-binding
model for the low-energy orbitals, with
Hamiltonian \begin{equation}\label{LRHami} H_{LR} = \Delta \epsilon (
  a_R^\dagger a_R - a_L^\dagger a_L) - t(a_R^\dagger a_L + a_L^
  \dagger a_R) + H_{LR(NL)},
\end{equation} where $a_L$ and $a_R$ are the annihilation operators for polaritons in orbitals localized on the left and right. $\Delta\epsilon$ is the detuning, i.e., the energy difference between these orbitals, and $t$ is the tunneling matrix element. We use \begin{equation}
H_{LR(NL)} = g[(a_L^\dagger a_L)^2 + (a_R^\dagger
a_R)^2], \label{eq:11}
\end{equation} to model the repulsive interactions within the condensates. Note that we have assumed, for simplicity, that the direct overlap of the localized orbitals is small, and hence neglected any off-diagonal interaction terms in the localized basis. We have also assumed that the localized orbitals are the same size, so that the interaction strength is the same for each.

The first step is to diagonalize the quadratic part of the Hamiltonian, and
transform to the basis of single-particle eigenstates. This is accomplished
by the standard transformation \begin{equation}\label{eq:LRtransform}
 \begin{pmatrix} a_L \\ a_R \end{pmatrix} = \begin{pmatrix} \cos(\theta) & -\sin(\theta)\\ \sin(\theta) & 
\cos(\theta) \end{pmatrix} \begin{pmatrix} a_1 \\ a_2 \end{pmatrix},
\end{equation} with $\tan(2\theta) = t/\Delta \epsilon$, which gives 
\begin{equation}\label{systemHami}
H = \Delta E ( a_2^\dagger a_2 - a_1^\dagger a_1) + H_{(NL)},
\end{equation} where $\Delta E=\sqrt{\Delta\epsilon^2+t^2}$. At zero detuning,
$a_1$ and $a_2$ annihilate particles in symmetric or antisymmetric orbitals
extended over the double well. 

The normalized transition strengths between the condensates and the
reservoirs $\alpha_{ij}$, defined in Eq.\,(\ref{eq:ntstrengths}),
follow from Eq.\,(\ref{eq:LRtransform}). Since the two reservoirs are
equivalent, these ratios are just the fraction of the density in each
orbital that lies over each reservoir, i.e., on the left or the right
of the junction:
\begin{align}
\alpha_{11} = \alpha_{22} = \cos^2(\theta) = \tfrac{1}{2} + \frac{\Delta \epsilon}{2 \sqrt{\Delta 
\epsilon^2 + t^2}} \nonumber \\
\alpha_{12} = \alpha_{21} = \sin^2(\theta) = \tfrac{1}{2} - \frac{\Delta \epsilon}{2 \sqrt{\Delta 
\epsilon^2 + t^2}}. \label{dwratios}
\end{align} Finally, we need the coefficients in the interaction Hamiltonian, Eq.\,(\ref{HNL}).
Writing Eq.\,(\ref{eq:11}) in the  $a_{1,2}$ basis, and comparing with Eq.\,(\ref{HNL}), gives
\begin{align}
&\kappa = g[\sin^4(\theta) + \cos^4(\theta)] = \frac{g}{2} \left(1 +
  \frac{\Delta \epsilon^2}{\Delta \epsilon^2 + t^2}\right), \nonumber \\ &\eta =
  8g\sin^2(\theta)\cos^2(\theta) = \frac{2 g \, t^2}{\Delta \epsilon^2
    + t^2}. \label{dwints}
\end{align} The transformation of Eq.\,(\ref{eq:11}) to the single-particle
eigenbasis also generates the interaction terms
\begin{equation}\begin{split} \frac{g}{2}\Big[&\sin^2(2\theta)a_1^\dagger a_1^\dagger a_2 a_2 \\
 & + \sin(4\theta)(a_2^\dagger a_2^\dagger a_2 a_1 - a_1^\dagger a_1^\dagger a_1 a_2)\Big] + \mathrm{h.c.},\label{eq:twowellj}
\end{split}\end{equation} which describe the scattering of particles between different orbitals. As discussed in Sec.\ \ref{sec:lowenh}, we
treat the quasilinear regime, where the strength of these terms is
smaller than the level spacing of the non-interacting Hamiltonian, so
that they are a small perturbation.  For the two-well problem the
condition for this approximation to be valid is parametrically
\begin{equation} \Delta E=\sqrt{\Delta\epsilon^2+t^2} \gtrsim g
  n.\label{eq:strapdefn}\end{equation} Here we neglect numerical
factors, which are of order 1 in typical geometries, including the
trigonometric functions in Eq. (\ref{eq:twowellj}). When
Eq. (\ref{eq:twowellj}) is satisfied the density generically undergoes
sinusoidal oscillations, while the relative phase between the
condensates can either oscillate or wind (linear ``Rabi'' oscillations
and linear ``Josephson'' oscillations,
respectively\cite{BlochJoseph2013}). For the special case $\Delta\epsilon=0$
Eq. (\ref{eq:twowellj}) is the criterion defining the Rabi regime as
given by Leggett.\cite{Leggett2001} It excludes any situation in which
the density imbalance is determined by interactions, rather than by
the pumping alone. We discuss this further in
Appendix~\ref{sec:regim-josephs-junct}.

\subsection{Population distribution}
\label{sec:popul-distr}
\begin{figure}[ht]
\includegraphics[width=8.5cm]{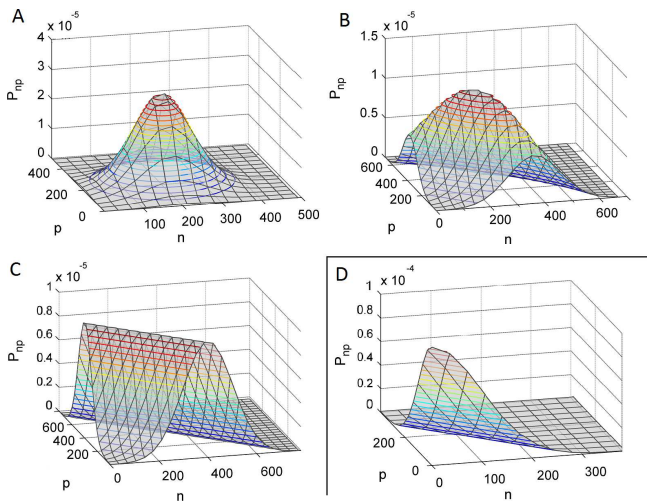}
\caption{\label{fig:populations} (Color online) Probability distribution of the
  populations $n,p$ of the two orbitals of a double-well potential,
  pumped by independent reservoirs for each well. Panels A--C: Results
  above the mean-field threshold, with $n_c=5200, \,n_s=5000$, and the
  reservoir to mode couplings varying as $\alpha_{11}=\alpha_{22}= 1$
  (A), 0.66 (B), 0.5 (C), and $\alpha_{12}=\alpha_{21} = 0$ (A), 0.33
  (B), 0.5 (C). This corresponds to decreasing the detuning or
  increasing the tunneling, so that the eigenstates evolve from
  localized orbitals of each well to delocalized,
  symmetric-antisymmetric superpositions.  A soft mode develops along
  the $n+p$ direction as the tunneling is increased and the modes
  become pumped indiscriminately by the two reservoirs.  Panel D:
  distribution at the mean-field threshold, $n_c=n_s=5000$, for
  $\alpha_{11}=\alpha_ {22}= 1$, $\alpha_{12}=\alpha_{21}= 0$.}
\end{figure}

In Fig.\,\ref{fig:populations} we show the population distribution
among the orbitals of the double well, obtained by using
Eq.\,(\ref{dwratios}) in Eq.\,(\ref{eq:10}). We take
$n_s$\,$=$\,$5000$, which is physically reasonable\cite{Whittaker2009}
while also in a regime which is convenient to contrast the approaches
presented in Sec.~\ref{sec:coherencefun}. We show results when the
pump parameter $n_c$ is both at the mean-field threshold
($n_c$\,$=$\,$n_s$; panel D), and above threshold ($n_c$\,$=$\,$5200$;
panels A--C). Above threshold we show how the distribution varies with
tunneling or detuning, giving results for three steps from vanishing
tunneling ($t/\Delta\epsilon$\,$=$\,$0$, panel A) to vanishing
detuning ($t/\Delta\epsilon$\,$\to$\,$\infty$, panel C). For vanishing
tunneling the orbitals are localized in the left and right well, and
the double-well system comprises two independent condensates. The
population distribution above threshold is then a symmetrical
two-dimensional Gaussian ($P_{np}$\,$=$\,$P_n P_p$).  However, as we
increase the tunneling, or decrease the detuning, the orbitals become
more delocalized between the left and right, and hence receive pumping
from both reservoirs. The distribution broadens along the direction
$n$\,$=$\,$p$, until at resonance the distribution is a flat ridge
along this direction ($P_{np}$\,$=$\,$P_{[n+p]}|_{n,p \ge 0}$).  This
is because the pumping is related to the density profile of the
condensate orbitals, which is identical for the two modes at
resonance. Thus, the pumping does not distinguish between the two
orbitals, fixing only the total density $n$\,$+$\,$p$, and leaving the
difference $n$\,$-$\,$p$ undetermined, within $n,p$\,$\ge$\,$0$.  In
other words, there is a soft mode describing density fluctuations
between the two condensate orbitals. The effects of the soft mode are
limited by the cut-off at $n$\,$=$\,$p$\,$=$\,$0$, so that the
population distribution in this case is not a Gaussian. Furthermore,
there are large fluctuations in the populations, which persist even
far above thres\-hold, $n_c$\,$\gg$\,$n_s$. We note that
$\Delta\epsilon$\,$=$\,$0$ corresponds to a nonequilibrium phase
boundary at the mean-field (rate-equation) level,\cite{Eastham2008}
separating the single-mode and two-mode steady-states. The large
fluctuations found here correspond to critical fluctuations near this
phase transition, associated with the finite size of the
condensate.\cite{eastham2006}

\subsection{First-order coherence functions}

\begin{figure}[ht!]
\includegraphics[width=8.7cm]{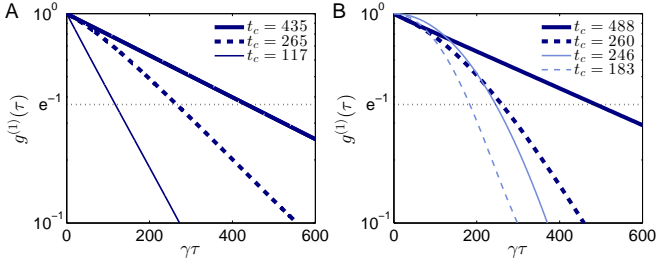}
\caption{\label{fig:linewidth} (Color online) First-order coherence functions for the
  emission from one mode of a double-well potential, with $g=
  4\times10^{-5}\gamma$ and $n_s=5000$. Solid curves show results for
  vanishing interwell tunneling ($t/\Delta \epsilon=0$), and dotted
  curves for vanishing detuning ($t/\Delta \epsilon\to\infty$).  Left
  panel (A): Numerical results with $n_c = 5000$ (thin line) and
  $5200$ (thick lines). Right panel (B): Static limit approximation
  (light blue) and Fokker-Planck approximation (dark blue), for $n_c =
  5200$. The insets give the corresponding coherence times in units of
  $1/\gamma$.}
\end{figure}

In Fig.\,\ref{fig:linewidth} we show the first-order coherence
functions for the light emitted from one mode of a double-well. The
left panel (A) shows the results obtained by direct numerical solution
of Eq.\,(\ref{linewidtheq}), while the right panel (B) shows those of
the Fokker-Planck and static-limit approaches. We again show results
at the mean-field threshold, and slightly above it, and for both
vanishing tunneling and vanishing detuning.
For all parameters shown, the numerical solution gives an exponential
decay of the first-order coherence, i.e., a Lorentzian emission
line. This is the same form found previously for a single
mode,\cite{Whittaker2009} and corresponds to interaction broadening in
the motional narrowing regime.\cite{Kubo1954, Hamm2005} That the
result is due to interaction effects is shown by the observation that
the computed coherence times are similar to those obtained from the
static limit calculation. The static limit, however, is deprived of
the motional narrowing effect\cite{Kubo1954, Whittaker2009} and shows
its hallmark Gaussian decay for the first-order coherence. For larger
polariton-polariton interaction strengths or broader population
distributions, the static regime dominates, and the numerical solution
and the static limit result coincide.

\begin{figure}[ht]
\includegraphics[width=8.7cm]{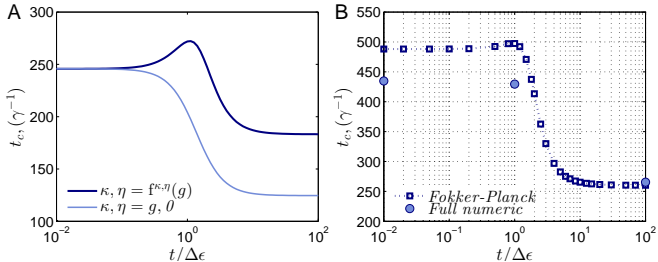} 
\caption{\label{fig:tc_effect} (Color online) Dependence of the coherence time (in
  units of $1/\gamma$) on the ratio of tunneling to detuning,
  $t/\Delta\epsilon$, for $n_c=5200$ and $n_s=5000$. Left panel (A):
  Static limit approximation, with $\kappa, \eta$ given by
  Eq.\,(\ref{dwints}) (dark blue), and with $\kappa, \eta = g,0$
  (light blue). Right panel (B): Fokker-Planck approximation and full
  numerical solution.}
\end{figure}

The full dependence of the coherence time, as predicted by the static
limit and Fokker-Planck approaches, is shown in
Fig.\,\ref{fig:tc_effect}. The coherence time decreases, so the
linewidth broadens, as we go from localized independent condensates
(zero tunneling) to delocalized coupled condensates (zero
detuning). This is because the broadening reflects the population
fluctuations, which cause energetic fluctuations via the interaction,
and the population fluctuations are largest for the delocalized case
(see Fig.\,\ref{fig:populations}). This effect is counterbalanced by
the changes in interaction strengths, $\kappa,\eta$, with detuning
Eqs.\,(\ref{dwints}).  Even though in the distributed case the
condensates overlap with one another and the average interaction
strength is stronger than in the independent case
($\frac{\kappa}{g},\frac{\eta}{g}$\,$=$\,$0.5, 2$ vs $1,0$), the
populations are anticorrelated between the modes. The resulting energy
fluctuations, therefore, to some extent cancel. To separate this
effect we show the coherence time obtained in the static limit taking
$\kappa$\,$=$\,$g$ and $\eta$\,$=$\,$0$ (panel A, light blue).  This
produces a basic step-like form of the coherence time.

When the dependence of the interaction strengths on detuning is
included two additional features are noticeable in the static limit
(panel A, dark blue): a non-monotonic dependence of the coherence
time, and an overall increase of the coherence time at $t/\Delta
\epsilon \approx 1$. The inter-condensate interaction is responsible
for these effects. The fluctuations in the populations of the modes
are anticorrelated at strong tunneling, because the total occupation,
$n\!+\!p$, is fixed by the pumping.  Since an increase in the
occupation of one mode tends to be accompanied by a decrease in the
occupation of the other, the energy shifts generated by the
interaction within the modes are partially canceled by those between
the modes. In this way the inter- and intra-condensate interactions
act in conjunction to preserve and even increase coherence. In the
motional narrowing regime, as shown in the Fokker-Planck and full
numerical solutions (panel B), fluctuations acquire a more isotropic
nature and this effect vanishes.

Finally, we see from Figs.\,\ref{fig:linewidth} and
\ref{fig:tc_effect} that the Fokker-Planck approximation is in good
agreement with the full numerical calculation. We note that, for
strong tunneling, the Fokker-Planck approximation indicates a Gaussian
behavior for the first-order coherence function, similar to that seen
in the static limit.  We suggest this is an effect of the soft density
mode. In the Fokker-Planck approach the drift coefficients are
appro\-ximated as linear functions of the populations, gi\-ving a
divergent relaxation time for fluctuations of $n$\,$-$\,$p$. Thus, the
soft mode has no dynamics at the level of the linearized theory, and
we obtain behavior similar to that of the static limit. In the full
theory, nonlinear effects are in place and this slow dynamics is
suppressed.

\subsection{Coherence of Josephson oscillations}

\begin{figure}[ht]
\includegraphics[width=8.5cm]{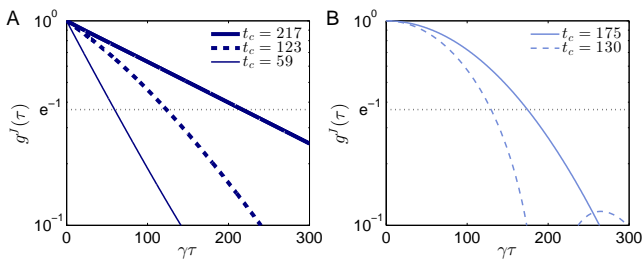}
\caption{\label{fig:Joseph} (Color online) Decay of the second-order cross-correlation
  function $g^J(\tau)$, describing the dephasing of intensity
  (Josephson) oscillations in the strong-trapping (Rabi) regime of a
  double-well potential. Corresponding results for the first-order coherence
  function are shown in Fig.~\ref{fig:linewidth}. Solid curves: Vanishing interwell
  tunneling. Dotted curves: vanishing detuning.  Left panel (A):
  Numerical results with $n_c = 5000$ (thin line), $5200$ (thick
  lines). Right panel (B): Static limit
  approximation for $n_c = 5200$. }
\end{figure}

In Fig.~\ref{fig:Joseph} we show the decay of the second-order
correlation function $g^{J}(\tau)$, describing the dephasing of the
intensity oscillations, obtained from both the numerical solution of
Eq. (\ref{eq:josephsonueq}), and from the static limit result,
Eq. (\ref{eq:josephsonstatic}). Comparing with
Fig.~\ref{fig:linewidth} we see that, when the interwell tunneling
vanishes, the decay time for $g^{J}(\tau)$ is half that of the
first-order coherence function $g^{(1)}(\tau)$, for the exponential
decays obtained from the numerical solution. For the Gaussian decays,
obtained in the static limit, the corresponding factor is
$1/\sqrt{2}$.  This is as expected, since for independent emitters
$g^{J}$ factorizes, $\langle a_1^\dagger(\tau)a_2(\tau)a_2^\dagger a_1
\rangle = \langle a_1^\dagger(\tau) a_1\rangle \langle a_2(\tau)
a_2^\dagger \rangle$. Such a factorization does not, in general, apply
in the case where the modes interact, and indeed we see that for the
numerical solution shown the decay of $g^{J}$ occurs faster than would
be expected from independent emitters.  This reflects the correlations
between the modes, which can be caused both by the Hamiltonian and the
dissipative interactions through the common reservoirs.

We make two additional comments regarding the relation between $g^{J}$
and $g^{(1)}$. First, for all the parameters shown the ratio of the
coherence times in the static limit appears to be $1/\sqrt{2}$, even
when the intermode interactions are present. One can readily check,
however, that $g^{J}$ does not factorize in this case, although this
is not apparent from these coherence times. Second, we note that for
independent emitters $g^{J}$ strictly factorizes into a product of
$g^{(1)}(\tau)$ and an antinormal ordered correlation function
$\langle a(\tau) a^\dagger\rangle$; the latter corresponds to an absorption
spectrum. The decay times for these two types of correlators can, in
principle, differ in an interacting system, where the dynamics after
adding a particle is not equivalent to that after removing one.  This
difference is negligible here, however, since the range of occupation
numbers in the steady-state density matrix is much larger than 1.

\section{\label{sec:discussion} Discussion}

Although we have focused on the specific example of condensation in a
double-well potential, our theory can be applied to a range of
many-condensate systems now being
considered,\cite{Lagoudakis2010,MasumotoYamamoto2012Bloch,TosiBaumberg2012NatPhys,GalbiatiBloch2012}
once they have been decomposed into the appropriate single-particle
orbitals. Each such orbital comprises a possible condensate mode in
our theory, with gain and loss characterized by a few phenomenological
parameters. In general the decoherence of the condensate depends on the
structure of the single particle orbitals, so our theory allows for
the study and optimization of coherence properties of polariton
condensates across the geometries now being developed, including
wires, photonic molecules, and photonic crystals. More speculatively,
it could provide a basis for studying glasslike
states\cite{Janot:2013,Malpuech:2007} and spontaneous vortex
lattices,\cite{Keeling2008} beyond the mean-field level, and also for
treatments of polariton dynamics in the quantum correlated
regime.\cite{carusottrev:2013}

One notable feature in experiments is the presence of a large
repulsive interaction with the reservoir
excitons.\cite{TosiBaumberg2012NatPhys,Wouters2:2008} We have omitted
this from our discussion, because it can be included on average as an
effective potential, and hence a redefinition of the
orbitals. Fluctuations in the reservoir occupation can broaden the
emission line, i.e., lead to decoherence of the condensate, but this
is negligible compared with the intrinsic linewidth provided pump
laser noise is small and fast.\cite{Love2008,Poddubny2013} Schwendimann, Quattropani,
and Sarchi have recently discussed an additional decoherence mechanism
for polariton condensates, involving parametric scattering processes,
and predicted its effects for a single-mode
condensate.\cite{SchwendimannQuattropaniSarchi2010} It would be
interesting to extend our theory to include this process, and hence
assess its impact in multimode condensates.

We have presented the theory without explicit consideration of the
polarization of the polaritons.\cite{ShelykhMalpuech2010} In incoherently pumped systems, the effects
of this additional degree of freedom have been explored in both
experiment and theory; the key result is that the condensate shows a
high degree of linear polarization, in a direction pinned to the
crystal
axes.\cite{Kasprzak2007,Klopotowski2006,Shelykh2006,Laussy2006} In
principle our theory allows the treatment of polarization, beyond the
mean-field (Gross-Pitaevskii) level, in two regimes.

Polarization can be included relatively straightforwardly within our
theory, when the polarization splitting of the single-particle
orbitals is negligible. In CdTe microcavities extrinsic effects,
probably strain, do induce such splittings, but they are typically
small, $\Omega_{ph}\sim \mu\mathrm{eV}$.\cite{Krizhanovskii2009}
Neglecting this scale we may take each spatial orbital in our theory
to comprise two degenerate circularly polarized orbitals, for
polaritons with $J_z=\pm 1$. Within each such orbital the interaction
Hamiltonian is\cite{ShelykhMalpuech2010} \begin{equation} c_1
  (n_+^2+n_-^2) + c_2 n_+n_-,\label{eq:polintn}\end{equation} where
$n_{+(-)}$ are the numbers of polaritons of each circular
polarization. Since this is of the same form as Eq.~(\ref{HNL}), i.e.,
diagonal in Fock space, in the circular basis, the form of our theory
is unaffected, although the number of modes is, in general,
doubled. 
(However, $|c_1|\gg |c_2|\approx
0$,\cite{MagnussonShelykh2010} so that if the polarizations have
independent reservoirs they completely decouple to a good
approximation, and the number of modes is effectively unchanged.)

Although a full analysis of this case is beyond the scope of this
paper, we can anticipate some results and ramifications. Each spatial
mode will give rise to two coherent circularly polarized emitters,
each with a linewidth determined by the co-polarized interactions (the
broadening due to the cross-polarized interactions will be smaller,
because $|c_2| \ll |c_1|$). The relative
occupation of the two circular polarizations will depend on the scattering
processes. In incoherently pumped systems, a reasonable assumption\cite{ShelykhMalpuech2010, Martin2002} is
that relaxation processes provide for the gain and nonlinear gain 
to be equivalent for the two circular polarizations, so that the resulting
occupations are identical, and the emission for each spatial mode is
then linearly polarized on average. This argument is very similar to
those previously used to explain the linear polarization of polariton
condensates:\cite{Shelykh2006,Laussy2006} the choice of the circular
basis arises because this diagonalizes the interaction Hamiltonian,
and the reasonable assumption of an equal population of such
eigenstates (which gives the minimum energy in equilibrium) is then a
linear polarization. If the two circular polarizations are truly
degenerate, $\Omega_{ph}=0$, then the direction of linear polarization
would fluctuate from shot to shot, but it can be locked by a non-zero
$\Omega_{ph}$.\cite{read2010} For a single spatial mode Laussy \emph{et
al}.\cite{Laussy2006} have gone beyond mean-field theory to predict the
decay time of the polarization, considering the Hamiltonian evolution
alone, i.e., in the static limit (c.f.
Sec~\ref{sec:first-order-coher}); we expect motional narrowing to
extend this decay time in general. 

The approach in terms of circularly polarized basis states breaks down
when, sufficiently far above threshold, the linewidths become smaller
than $\Omega_{ph}$, which is no longer a small energy scale. In this
regime one should take the (typically) linearly polarized eigenstates
of the single-particle Hamiltonian as the starting point. However,
transforming the interaction, Eq. (\ref{eq:polintn}), to this basis
leads to polarization-flip scattering terms, such as $a_x^\dagger
a_x^\dagger a_y a_y$. This is of the form of
Eq. (\ref{eq:gentwomodes}), but unfortunately since $\Omega_{ph}$ is
usually small compared with the interaction energy it cannot be
similarly neglected. Thus this regime cannot be fully treated within
the framework of our theory, unless it is extended to include
non-conserving scattering processes, i.e., spin-flip terms. An
exception is when only one of the two orthogonal polarizations is
populated, which can indeed occur,\cite{Krizhanovskii2009} so that the
unoccupied orthogonally polarized mode may be omitted completely from
the description.

\section{\label{sec:conclusion} Conclusion}

In summary, we have developed a model for the nonequilibrium dynamics
of polaritons in an incoherently pumped microcavity, incorporating
gain due to scattering from multiple reservoirs, and resonant
polariton-polariton interactions. In contrast to previous works
addressing condensates formed with a single macroscopically-occupied
orbital, our theory applies when several such states coexist, i.e., to
multimode polariton condensates. We have used it to predict the
quantum statistics, revealed for example via the linewidths, and shown
how these quantities are affected by interactions between the
condensates. We predict that the populations of the modes can be
anticorrelated due to their coupling to a common reservoir, leading to
a narrowing of the emission lines and a prolongation of the coherence
time. We have also demonstrated theoretically a dephasing mechanism
for intensity oscillations, and shown that, for realistic parameters,
their coherence decay provides a useful probe of correlation effects.

An important theoretical extension of our work would be to include the
nonresonant interaction terms between the modes, in particular terms
such as $a_1^\dagger a_1^\dagger a_2 a_2$, which become significant
beyond the strong-trapping regime. This would also allow us to include spin-flip
and polarization dependent mechanisms.\cite{GlazovKavokin2013,MagnussonShelykh2010}
In Fock space, the Liouville evolution of these terms generates a recursive
dependence on all the elements within density matrix. This contrasts with having
to solve for the diagonal elements in the case of steady-state population
distribution or the one off-diagonal terms for linewidth and Josephson
coherence function. We suggest that these interactions could be
included by generalizing the Fokker-Planck approach to apply to the
full density operator, rather than the distributions $u$ or $P$, i.e.,
by assuming $\rho_{mnpq}$ is smooth, so that Eq.~(\ref{eq:1}) becomes
a partial differential equation. Such an approach would be similar in
spirit to those based on the Wigner representation for $\rho$, as
discussed by Wouters and Savona\cite{wouters09} among others. It could
also be done without consideration of the classical limit of quantum electrodynamics
and make use of the Mellin transform in relation to fractional calculus,\cite{FracCal2007,FracCal2011}
in contrast to the double-sided Laplace transform which led to Eq.~(\ref{eq:g1result}).
More numerically driven, the cumulant expansion technique used in
Ref. \onlinecite{MagnussonShelykh2010} may also lead to a way to deal
with these terms. The non resonant interaction terms lead to the Bogoliubov
spectrum for a homogeneous single-mode condensate, and hence are implicated
in superfluidity, while at the semiclassical level they cause nonlinear mixing and
synchronization in the multimode case.\cite{Eastham2008} The suggested
generalization of our theory would allow the impact of quantum and
nonequilibrium fluctuations on such phenomena to be explored, in
complex geometries where many condensates coexist. Josephson phenomena which
occur outside the strong-trapping regime are also accessible once these
terms are included.

\acknowledgments

This work was supported by Science Foundation Ireland
(09/SIRG/I1592). Complementary funding was obtained through
the POLATOM Network of the European Science Foundation. Discussions with
D. Whittaker are acknowledged.


\appendix

\section{Derivation of the Effective Pumping Rate}\label{sec:appxA}

To derive an effective pumping rate, $r$, starting from the replenishing rate, $\lambda_a$, of reservoir level $|a\rangle$, we combine the system to a generic level, $\Phi=\rho^{\phi\phi}|\phi\rangle\!\langle\phi|$, giving $\rho \otimes \Phi$. The evolution of the coupled system, projected onto $|\phi\rangle$, provides the rate equation,
\begin{equation}
 0 = -\lambda_a \rho_{nmpq}^{\phi\phi} +\gamma_r(\rho_{nmpq}^{aa}+\rho_{nmpq}^{b_1b_1}+\rho_{nmpq}^{b_2b_2}), \label{eq:solvedphi}
\end{equation}
which we use in conjunction with the trace over all levels, $\mathrm{Tr}_{a,b_i,\phi}[\rho \otimes \Phi]$,
\begin{equation}
 \rho_{nmpq}= \rho_{nmpq}^{\phi\phi} + \rho_{nmpq}^{aa} + \rho_{nmpq}^{b_1b_1} + \rho_{nmpq}^{b_2b_2} \label{eq:aabb's}
\end{equation}
Substituting $\rho_{nmpq}-\rho_{nmpq}^{\phi\phi}$ for $\rho_{nmpq}^{aa}+\rho_{nmpq}^{b_1b_1}+\rho_{nmpq}^{b_2b_2}$, (\ref{eq:aabb's}), into (\ref{eq:solvedphi}) gives
\begin{equation}
 \rho_{nmpq}^{\phi\phi} = \frac{\gamma_r}{\lambda_a+\gamma_r}\rho_{nmpq}.
\end{equation}
Hence
\begin{equation}
 \lambda_a \rho_{nmpq}^{\phi\phi} = \cfrac{\lambda_a \gamma_r}{\lambda_a + \gamma_r} \rho_{nmpq} = r \rho_{nmpq},
\end{equation}
and we have transformed the repopulation from level $\Phi$ into an effective pumping, with rate $r$. Additional levels could be included, but their steady-state nature allows us to recover this simpler scheme.\cite{Scully&Zubairy1997}

\section{Intermediate Vector and Matrix Forms in the Simplified Pumping Model}\label{sec:appxB}

To obtain the one reservoir, two modes dissipator (\ref{eq:4}), we generate three versions of these vector forms,
\begin{equation}
R=
\begin{pmatrix}
\rho_{nmpq}^{aa}\\
\rho_{nm+1pq}^{ab_1}\\
\rho_{n+1mpq}^{b_1a}\\
\rho_{nmpq+1}^{ab_2}\\
\rho_{nmp+1q}^{b_2a}\\
\rho_{n+1mpq+1}^{b_1b_2}\\
\rho_{nm+1p+1q}^{b_2b_1}\\
\rho_{n+1m+1pq}^{b_1b_1}\\
\rho_{nmp+1q+1}^{b_2b_2}
\end{pmatrix}, \quad
A=r 
\begin{pmatrix}
 \rho_{nmpq}\\ 0\\ 0\\ 0\\ 0\\ 0\\ 0\\ 0\\ 0
\end{pmatrix}.
\end{equation}
\begin{widetext}
\hspace{-4mm}We also use three instances of the following matrix form,
\begin{scriptsize}
\begin{equation}
M=
\begin{pmatrix} \label{matrix_M}
\gamma_r &-ig_1\sqrt{m+1} &ig_1\sqrt{n+1} &-ig_2\sqrt{q+1} &ig_2\sqrt{p+1} &0 &0 &0 &0\\
-ig_1\sqrt{m+1} &\gamma_r &0 &0 &0 &0 &ig_2\sqrt{p+1} &ig_1\sqrt{n+1} &0\\
ig_1\sqrt{n+1} &0 &\gamma_r &0 &0 &-ig_2\sqrt{q+1} &0 &-ig_1\sqrt{m+1} &0\\
-ig_2\sqrt{q+1} &0 &0 &\gamma_r &0 &ig_1\sqrt{n+1} &0 &0 &ig_2\sqrt{p+1}\\
ig_2\sqrt{p+1} &0 &0 &0 &\gamma_r &0 &-ig_1\sqrt{m+1} &0 &-ig_2\sqrt{q+1}\\
0 &0 &-ig_2\sqrt{q+1} &ig_1\sqrt{n+1} &0 &\gamma_r &0 &0 &0\\
0 &ig_2\sqrt{p+1} &0 &0 &-ig_1\sqrt{m+1} &0 &\gamma_r &0 &0\\
0 &ig_1\sqrt{n+1} &-ig_1\sqrt{m+1} &0 &0 &0 &0 &\gamma_r &0\\
0 &0 &0 &ig_2\sqrt{p+1} &-ig_2\sqrt{q+1} &0 &0 &0 &\gamma_r\\
\end{pmatrix}.
\end{equation}
\end{scriptsize}
The vectors and matrices $R',A',M'$ and $R'',A'',M''$ are obtained by shifting the indices and occupation numbers of $R,A,M$ according to $n,m \rightarrow n-\!1,m-\!1$ and $p,q \rightarrow p\!-\!1, q\!-\!1$, respectively. We use the elements $M_{21}^{-1}..M_{51}^{-1}$, $M'^{-1}_{21},M'^{-1}_{31}$ and $M''^{-1}_{41},M''^{-1}_{51}$ in our substitution. 
\end{widetext}

\section{Regimes of Josephson junctions}

\label{sec:regim-josephs-junct}

Within the mean-field (Gross-Pitaevskii) dynamics of a Josephson
junction one typically discusses several different regimes, and a
range of Josephson effects within each regime. If
Eq. (\ref{eq:strapdefn}) holds then the mean-field dynamics is that of
two coupled harmonic oscillators, leading to our description of this
regime as linear. In it one obtains sinusoidal oscillations in the
density from the beating between normal modes.\cite{javanainen1986}
Since this is also the physics of Rabi oscillations of a two-state
system, the linear regime is sometimes also described as the Rabi
regime.  In the literature the criterion for the linear/Rabi
regime\cite{Leggett2001,BECJosephTheory,BlochJoseph2013} is usually
stated for the case where the detuning, $\Delta\epsilon$, is zero, or
at least similar to the tunneling, $t$, and so is $t\gtrsim gn$. For
$\Delta\epsilon\neq 0$ the behavior at the mean-field level can remain
linear (constant blueshifts, of order $gn$, excepted) even for very
small tunneling, so long as Eq. (\ref{eq:strapdefn}) is satisfied. The
opposite regime, where the interactions dominate, is usually labeled
as the Josephson regime, and is where macroscopic quantum
self-trapping is studied.

The experiment involving continuous incoherent
pumping,\cite{Lagoudakis2010} as in our theory, is generally agreed to
be in the linear/Rabi regime,\cite{Lagoudakis2010,BlochJoseph2013} so
that we expect our theory to apply. We note, however, one potential
complication. If interactions like Eq.~(\ref{eq:twowellj}) are
completely neglected then there are no terms which fix the relative
phase of the condensates in the two modes. Thus the phase of the
Josephson oscillations would fluctuate from shot to shot of an
experiment [see Eq. (\ref{eq:densosc})]. The data reported in Ref.\
\onlinecite{Lagoudakis2010} are, however, averaged over many
repetitions and still reveal oscillations, so that a consistent phase
is being established. Thus the terms in Eq.~(\ref{eq:twowellj}) do
have some effect, perhaps implying some small corrections to our
results.

As noted above, some other recent experiments involve coherent
resonant excitation to create a transient
condensate.\cite{BlochJoseph2013} The linear regime discussed by these
authors would correspond to Eq. (\ref{eq:strapdefn}) being satisfied,
while the nonlinear regime (where macroscopic quantum self-trapping
was observed) would be where it is violated; the further distinction
made within the linear regime, between ``Rabi'' and ``Josephson''
oscillations, would not be relevant in terms of the applicability of
our theory. It is, in any case, not directly relevant to these
experiments, as they have neither a condensed steady-state nor
incoherent excitation.


\newpage


\begin{thebibliography}{69}%
\makeatletter
\providecommand \@ifxundefined [1]{%
 \@ifx{#1\undefined}
}%
\providecommand \@ifnum [1]{%
 \ifnum #1\expandafter \@firstoftwo
 \else \expandafter \@secondoftwo
 \fi
}%
\providecommand \@ifx [1]{%
 \ifx #1\expandafter \@firstoftwo
 \else \expandafter \@secondoftwo
 \fi
}%
\providecommand \natexlab [1]{#1}%
\providecommand \enquote  [1]{``#1''}%
\providecommand \bibnamefont  [1]{#1}%
\providecommand \bibfnamefont [1]{#1}%
\providecommand \citenamefont [1]{#1}%
\providecommand \href@noop [0]{\@secondoftwo}%
\providecommand \href [0]{\begingroup \@sanitize@url \@href}%
\providecommand \@href[1]{\@@startlink{#1}\@@href}%
\providecommand \@@href[1]{\endgroup#1\@@endlink}%
\providecommand \@sanitize@url [0]{\catcode `\\12\catcode `\$12\catcode
  `\&12\catcode `\#12\catcode `\^12\catcode `\_12\catcode `\%12\relax}%
\providecommand \@@startlink[1]{}%
\providecommand \@@endlink[0]{}%
\providecommand \url  [0]{\begingroup\@sanitize@url \@url }%
\providecommand \@url [1]{\endgroup\@href {#1}{\urlprefix }}%
\providecommand \urlprefix  [0]{URL }%
\providecommand \Eprint [0]{\href }%
\providecommand \doibase [0]{http://dx.doi.org/}%
\providecommand \selectlanguage [0]{\@gobble}%
\providecommand \bibinfo  [0]{\@secondoftwo}%
\providecommand \bibfield  [0]{\@secondoftwo}%
\providecommand \translation [1]{[#1]}%
\providecommand \BibitemOpen [0]{}%
\providecommand \bibitemStop [0]{}%
\providecommand \bibitemNoStop [0]{.\EOS\space}%
\providecommand \EOS [0]{\spacefactor3000\relax}%
\providecommand \BibitemShut  [1]{\csname bibitem#1\endcsname}%
\let\auto@bib@innerbib\@empty
\bibitem [{\citenamefont {Hopfield}(1958)}]{Hopfield1958}%
  \BibitemOpen
  \bibfield  {author} {\bibinfo {author} {\bibfnamefont {J.~J.}\ \bibnamefont
  {Hopfield}},\ }\href@noop {} {\bibfield  {journal} {\bibinfo  {journal}
  {Phys. Rev.}\ }\textbf {\bibinfo {volume} {112}},\ \bibinfo {pages} {1555}
  (\bibinfo {year} {1958})}\BibitemShut {NoStop}%
\bibitem [{\citenamefont {Weisbuch}\ \emph {et~al.}(1992)\citenamefont
  {Weisbuch}, \citenamefont {Nishioka}, \citenamefont {Ishikawa},\ and\
  \citenamefont {Arakawa}}]{Weisbuch1992}%
  \BibitemOpen
  \bibfield  {author} {\bibinfo {author} {\bibfnamefont {C.}~\bibnamefont
  {Weisbuch}}, \bibinfo {author} {\bibfnamefont {M.}~\bibnamefont {Nishioka}},
  \bibinfo {author} {\bibfnamefont {A.}~\bibnamefont {Ishikawa}}, \ and\
  \bibinfo {author} {\bibfnamefont {Y.}~\bibnamefont {Arakawa}},\ }\href@noop
  {} {\bibfield  {journal} {\bibinfo  {journal} {Phys. Rev. Lett.}\ }\textbf
  {\bibinfo {volume} {69}},\ \bibinfo {pages} {3314} (\bibinfo {year}
  {1992})}\BibitemShut {NoStop}%
\bibitem [{\citenamefont {Dang}\ \emph {et~al.}(1998)\citenamefont {Dang},
  \citenamefont {Heger}, \citenamefont {Andr{\'e}}, \citenamefont {Boeuf},\
  and\ \citenamefont {Romestain}}]{Dang1998}%
  \BibitemOpen
  \bibfield  {author} {\bibinfo {author} {\bibfnamefont {L.~S.}\ \bibnamefont
  {Dang}}, \bibinfo {author} {\bibfnamefont {D.}~\bibnamefont {Heger}},
  \bibinfo {author} {\bibfnamefont {R.}~\bibnamefont {Andr{\'e}}}, \bibinfo
  {author} {\bibfnamefont {F.}~\bibnamefont {Boeuf}}, \ and\ \bibinfo {author}
  {\bibfnamefont {R.}~\bibnamefont {Romestain}},\ }\href@noop {} {\bibfield
  {journal} {\bibinfo  {journal} {Phys. Rev. Lett.}\ }\textbf {\bibinfo
  {volume} {81}},\ \bibinfo {pages} {3920} (\bibinfo {year}
  {1998})}\BibitemShut {NoStop}%
\bibitem [{\citenamefont {Kasprzak}\ \emph {et~al.}(2006)\citenamefont
  {Kasprzak}, \citenamefont {Richard}, \citenamefont {Kundermann},
  \citenamefont {Baas}, \citenamefont {Jeambrun}, \citenamefont {Keeling},
  \citenamefont {Marchetti}, \citenamefont {Szymanska}, \citenamefont {Andre},
  \citenamefont {Staehli}, \citenamefont {Savona}, \citenamefont {Littlewood},
  \citenamefont {Deveaud},\ and\ \citenamefont {Dang}}]{Kasprzak2006}%
  \BibitemOpen
  \bibfield  {author} {\bibinfo {author} {\bibfnamefont {J.}~\bibnamefont
  {Kasprzak}}, \bibinfo {author} {\bibfnamefont {M.}~\bibnamefont {Richard}},
  \bibinfo {author} {\bibfnamefont {S.}~\bibnamefont {Kundermann}}, \bibinfo
  {author} {\bibfnamefont {A.}~\bibnamefont {Baas}}, \bibinfo {author}
  {\bibfnamefont {P.}~\bibnamefont {Jeambrun}}, \bibinfo {author}
  {\bibfnamefont {J.~M.~J.}\ \bibnamefont {Keeling}}, \bibinfo {author}
  {\bibfnamefont {F.~M.}\ \bibnamefont {Marchetti}}, \bibinfo {author}
  {\bibfnamefont {M.~H.}\ \bibnamefont {Szymanska}}, \bibinfo {author}
  {\bibfnamefont {R.}~\bibnamefont {Andre}}, \bibinfo {author} {\bibfnamefont
  {J.~L.}\ \bibnamefont {Staehli}}, \bibinfo {author} {\bibfnamefont
  {V.}~\bibnamefont {Savona}}, \bibinfo {author} {\bibfnamefont {P.~B.}\
  \bibnamefont {Littlewood}}, \bibinfo {author} {\bibfnamefont
  {B.}~\bibnamefont {Deveaud}}, \ and\ \bibinfo {author} {\bibfnamefont
  {L.~S.}\ \bibnamefont {Dang}},\ }\href@noop {} {\bibfield  {journal}
  {\bibinfo  {journal} {Nature}\ }\textbf {\bibinfo {volume} {443}},\ \bibinfo
  {pages} {409} (\bibinfo {year} {2006})}\BibitemShut {NoStop}%
\bibitem [{\citenamefont {Love}\ \emph {et~al.}(2008)\citenamefont {Love},
  \citenamefont {Krizhanovskii}, \citenamefont {Whittaker}, \citenamefont
  {Bouchekioua}, \citenamefont {Sanvitto}, \citenamefont {Rizeiqi},
  \citenamefont {Bradley}, \citenamefont {Skolnick}, \citenamefont {Eastham},
  \citenamefont {Andr{\'e}},\ and\ \citenamefont {Dang}}]{Love2008}%
  \BibitemOpen
  \bibfield  {author} {\bibinfo {author} {\bibfnamefont {A.~P.~D.}\
  \bibnamefont {Love}}, \bibinfo {author} {\bibfnamefont {D.~N.}\ \bibnamefont
  {Krizhanovskii}}, \bibinfo {author} {\bibfnamefont {D.~M.}\ \bibnamefont
  {Whittaker}}, \bibinfo {author} {\bibfnamefont {R.}~\bibnamefont
  {Bouchekioua}}, \bibinfo {author} {\bibfnamefont {D.}~\bibnamefont
  {Sanvitto}}, \bibinfo {author} {\bibfnamefont {S.~A.}\ \bibnamefont
  {Rizeiqi}}, \bibinfo {author} {\bibfnamefont {R.}~\bibnamefont {Bradley}},
  \bibinfo {author} {\bibfnamefont {M.~S.}\ \bibnamefont {Skolnick}}, \bibinfo
  {author} {\bibfnamefont {P.~R.}\ \bibnamefont {Eastham}}, \bibinfo {author}
  {\bibfnamefont {R.}~\bibnamefont {Andr{\'e}}}, \ and\ \bibinfo {author}
  {\bibfnamefont {L.~S.}\ \bibnamefont {Dang}},\ }\href@noop {} {\bibfield
  {journal} {\bibinfo  {journal} {Phys. Rev. Lett.}\ }\textbf {\bibinfo
  {volume} {101}},\ \bibinfo {pages} {067404} (\bibinfo {year}
  {2008})}\BibitemShut {NoStop}%
\bibitem [{\citenamefont {Assmann}\ \emph {et~al.}(2011)\citenamefont
  {Assmann}, \citenamefont {Tempel}, \citenamefont {Veit}, \citenamefont
  {Bayer}, \citenamefont {Rahimi-Iman}, \citenamefont {Loeffler}, \citenamefont
  {Hoefling}, \citenamefont {Reitzenstein}, \citenamefont {Worschech},\ and\
  \citenamefont {Forchel}}]{AssmannForchel2011}%
  \BibitemOpen
  \bibfield  {author} {\bibinfo {author} {\bibfnamefont {M.}~\bibnamefont
  {Assmann}}, \bibinfo {author} {\bibfnamefont {J.-S.}\ \bibnamefont {Tempel}},
  \bibinfo {author} {\bibfnamefont {F.}~\bibnamefont {Veit}}, \bibinfo {author}
  {\bibfnamefont {M.}~\bibnamefont {Bayer}}, \bibinfo {author} {\bibfnamefont
  {A.}~\bibnamefont {Rahimi-Iman}}, \bibinfo {author} {\bibfnamefont
  {A.}~\bibnamefont {Loeffler}}, \bibinfo {author} {\bibfnamefont
  {S.}~\bibnamefont {Hoefling}}, \bibinfo {author} {\bibfnamefont
  {S.}~\bibnamefont {Reitzenstein}}, \bibinfo {author} {\bibfnamefont
  {L.}~\bibnamefont {Worschech}}, \ and\ \bibinfo {author} {\bibfnamefont
  {A.}~\bibnamefont {Forchel}},\ }\href@noop {} {\bibfield  {journal} {\bibinfo
   {journal} {Proc. Natl. Acad. Sci. USA}\ }\textbf {\bibinfo {volume} {108}},\
  \bibinfo {pages} {1804} (\bibinfo {year} {2011})}\BibitemShut {NoStop}%
\bibitem [{\citenamefont {Pitaevskii}\ and\ \citenamefont
  {Stringari}(2003)}]{Pitaevskii2003}%
  \BibitemOpen
  \bibfield  {author} {\bibinfo {author} {\bibfnamefont {L.}~\bibnamefont
  {Pitaevskii}}\ and\ \bibinfo {author} {\bibfnamefont {S.}~\bibnamefont
  {Stringari}},\ }\href@noop {} {\emph {\bibinfo {title} {{Bose-Einstein
  Condensation}}}},\ \bibinfo {series} {{International Series of Monographs on
  Physics}}, Vol.\ \bibinfo {volume} {116}\ (\bibinfo  {publisher} {Oxford
  University Press, London},\ \bibinfo {year} {2003})\BibitemShut {NoStop}%
\bibitem [{\citenamefont {Tosi}\ \emph {et~al.}(2012)\citenamefont {Tosi},
  \citenamefont {Christmann}, \citenamefont {Berloff}, \citenamefont {Tsotsis},
  \citenamefont {Gao}, \citenamefont {Hatzopoulos}, \citenamefont {Savvidis},\
  and\ \citenamefont {Baumberg}}]{TosiBaumberg2012NatPhys}%
  \BibitemOpen
  \bibfield  {author} {\bibinfo {author} {\bibfnamefont {G.}~\bibnamefont
  {Tosi}}, \bibinfo {author} {\bibfnamefont {G.}~\bibnamefont {Christmann}},
  \bibinfo {author} {\bibfnamefont {N.~G.}\ \bibnamefont {Berloff}}, \bibinfo
  {author} {\bibfnamefont {P.}~\bibnamefont {Tsotsis}}, \bibinfo {author}
  {\bibfnamefont {T.}~\bibnamefont {Gao}}, \bibinfo {author} {\bibfnamefont
  {Z.}~\bibnamefont {Hatzopoulos}}, \bibinfo {author} {\bibfnamefont {P.~G.}\
  \bibnamefont {Savvidis}}, \ and\ \bibinfo {author} {\bibfnamefont {J.~J.}\
  \bibnamefont {Baumberg}},\ }\href@noop {} {\bibfield  {journal} {\bibinfo
  {journal} {Nat. Phys.}\ }\textbf {\bibinfo {volume} {8}},\ \bibinfo {pages}
  {190} (\bibinfo {year} {2012})}\BibitemShut {NoStop}%
\bibitem [{\citenamefont {Galbiati}\ \emph {et~al.}(2012)\citenamefont
  {Galbiati}, \citenamefont {Ferrier}, \citenamefont {Solnyshkov},
  \citenamefont {Tanese}, \citenamefont {Wertz}, \citenamefont {Amo},
  \citenamefont {Abbarchi}, \citenamefont {Senellart}, \citenamefont {Sagnes},
  \citenamefont {Lema{\^i}tre}, \citenamefont {Galopin}, \citenamefont
  {Malpuech},\ and\ \citenamefont {Bloch}}]{GalbiatiBloch2012}%
  \BibitemOpen
  \bibfield  {author} {\bibinfo {author} {\bibfnamefont {M.}~\bibnamefont
  {Galbiati}}, \bibinfo {author} {\bibfnamefont {L.}~\bibnamefont {Ferrier}},
  \bibinfo {author} {\bibfnamefont {D.~D.}\ \bibnamefont {Solnyshkov}},
  \bibinfo {author} {\bibfnamefont {D.}~\bibnamefont {Tanese}}, \bibinfo
  {author} {\bibfnamefont {E.}~\bibnamefont {Wertz}}, \bibinfo {author}
  {\bibfnamefont {A.}~\bibnamefont {Amo}}, \bibinfo {author} {\bibfnamefont
  {M.}~\bibnamefont {Abbarchi}}, \bibinfo {author} {\bibfnamefont
  {P.}~\bibnamefont {Senellart}}, \bibinfo {author} {\bibfnamefont
  {I.}~\bibnamefont {Sagnes}}, \bibinfo {author} {\bibfnamefont
  {A.}~\bibnamefont {Lema{\^i}tre}}, \bibinfo {author} {\bibfnamefont
  {E.}~\bibnamefont {Galopin}}, \bibinfo {author} {\bibfnamefont
  {G.}~\bibnamefont {Malpuech}}, \ and\ \bibinfo {author} {\bibfnamefont
  {J.}~\bibnamefont {Bloch}},\ }\href@noop {} {\bibfield  {journal} {\bibinfo
  {journal} {Phys. Rev. Lett.}\ }\textbf {\bibinfo {volume} {108}},\ \bibinfo
  {pages} {126403} (\bibinfo {year} {2012})}\BibitemShut {NoStop}%
\bibitem [{\citenamefont {Lagoudakis}\ \emph {et~al.}(2010)\citenamefont
  {Lagoudakis}, \citenamefont {Pietka}, \citenamefont {Wouters}, \citenamefont
  {Andr{\'e}},\ and\ \citenamefont {Deveaud-Pl{\'e}dran}}]{Lagoudakis2010}%
  \BibitemOpen
  \bibfield  {author} {\bibinfo {author} {\bibfnamefont {K.~G.}\ \bibnamefont
  {Lagoudakis}}, \bibinfo {author} {\bibfnamefont {B.}~\bibnamefont {Pietka}},
  \bibinfo {author} {\bibfnamefont {M.}~\bibnamefont {Wouters}}, \bibinfo
  {author} {\bibfnamefont {R.}~\bibnamefont {Andr{\'e}}}, \ and\ \bibinfo
  {author} {\bibfnamefont {B.}~\bibnamefont {Deveaud-Pl{\'e}dran}},\
  }\href@noop {} {\bibfield  {journal} {\bibinfo  {journal} {Phys. Rev. Lett.}\
  }\textbf {\bibinfo {volume} {105}},\ \bibinfo {pages} {120403} (\bibinfo
  {year} {2010})}\BibitemShut {NoStop}%
\bibitem [{\citenamefont {Abbarchi}\ \emph {et~al.}(2013)\citenamefont
  {Abbarchi}, \citenamefont {Amo}, \citenamefont {Sala}, \citenamefont
  {Solnyshkov}, \citenamefont {Flayac}, \citenamefont {Ferrier}, \citenamefont
  {Sagnes}, \citenamefont {Galopin}, \citenamefont {Lema\^itre}, \citenamefont
  {Malpuech},\ and\ \citenamefont {Bloch}}]{BlochJoseph2013}%
  \BibitemOpen
  \bibfield  {author} {\bibinfo {author} {\bibfnamefont {M.}~\bibnamefont
  {Abbarchi}}, \bibinfo {author} {\bibfnamefont {A.}~\bibnamefont {Amo}},
  \bibinfo {author} {\bibfnamefont {V.~G.}\ \bibnamefont {Sala}}, \bibinfo
  {author} {\bibfnamefont {D.~D.}\ \bibnamefont {Solnyshkov}}, \bibinfo
  {author} {\bibfnamefont {H.}~\bibnamefont {Flayac}}, \bibinfo {author}
  {\bibfnamefont {L.}~\bibnamefont {Ferrier}}, \bibinfo {author} {\bibfnamefont
  {I.}~\bibnamefont {Sagnes}}, \bibinfo {author} {\bibfnamefont
  {E.}~\bibnamefont {Galopin}}, \bibinfo {author} {\bibfnamefont
  {A.}~\bibnamefont {Lema\^itre}}, \bibinfo {author} {\bibfnamefont
  {G.}~\bibnamefont {Malpuech}}, \ and\ \bibinfo {author} {\bibfnamefont
  {J.}~\bibnamefont {Bloch}},\ }\href@noop {} {\bibfield  {journal} {\bibinfo
  {journal} {Nat. Phys.}\ }\textbf {\bibinfo {volume} {9}},\ \bibinfo {pages}
  {275} (\bibinfo {year} {2013})}\BibitemShut {NoStop}%
\bibitem [{\citenamefont {Lai}\ \emph {et~al.}(2007)\citenamefont {Lai},
  \citenamefont {Kim}, \citenamefont {Utsunomiya}, \citenamefont {Roumpos},
  \citenamefont {Deng}, \citenamefont {Fraser}, \citenamefont {Byrnes},
  \citenamefont {Recher}, \citenamefont {Kumada}, \citenamefont {Fujisawa},\
  and\ \citenamefont {Yamamoto}}]{LaiYamamoto2007Nat}%
  \BibitemOpen
  \bibfield  {author} {\bibinfo {author} {\bibfnamefont {C.~W.}\ \bibnamefont
  {Lai}}, \bibinfo {author} {\bibfnamefont {N.~Y.}\ \bibnamefont {Kim}},
  \bibinfo {author} {\bibfnamefont {S.}~\bibnamefont {Utsunomiya}}, \bibinfo
  {author} {\bibfnamefont {G.}~\bibnamefont {Roumpos}}, \bibinfo {author}
  {\bibfnamefont {H.}~\bibnamefont {Deng}}, \bibinfo {author} {\bibfnamefont
  {M.~D.}\ \bibnamefont {Fraser}}, \bibinfo {author} {\bibfnamefont
  {T.}~\bibnamefont {Byrnes}}, \bibinfo {author} {\bibfnamefont
  {P.}~\bibnamefont {Recher}}, \bibinfo {author} {\bibfnamefont
  {N.}~\bibnamefont {Kumada}}, \bibinfo {author} {\bibfnamefont
  {T.}~\bibnamefont {Fujisawa}}, \ and\ \bibinfo {author} {\bibfnamefont
  {Y.}~\bibnamefont {Yamamoto}},\ }\href@noop {} {\bibfield  {journal}
  {\bibinfo  {journal} {Nature}\ }\textbf {\bibinfo {volume} {450}},\ \bibinfo
  {pages} {529} (\bibinfo {year} {2007})}\BibitemShut {NoStop}%
\bibitem [{\citenamefont {Masumoto}\ \emph {et~al.}(2012)\citenamefont
  {Masumoto}, \citenamefont {Kim}, \citenamefont {Byrnes}, \citenamefont
  {Kusudo}, \citenamefont {L{\"o}ffler}, \citenamefont {H{\"o}fling},
  \citenamefont {Forchel},\ and\ \citenamefont
  {Yamamoto}}]{MasumotoYamamoto2012Bloch}%
  \BibitemOpen
  \bibfield  {author} {\bibinfo {author} {\bibfnamefont {N.}~\bibnamefont
  {Masumoto}}, \bibinfo {author} {\bibfnamefont {N.~Y.}\ \bibnamefont {Kim}},
  \bibinfo {author} {\bibfnamefont {T.}~\bibnamefont {Byrnes}}, \bibinfo
  {author} {\bibfnamefont {K.}~\bibnamefont {Kusudo}}, \bibinfo {author}
  {\bibfnamefont {A.}~\bibnamefont {L{\"o}ffler}}, \bibinfo {author}
  {\bibfnamefont {S.}~\bibnamefont {H{\"o}fling}}, \bibinfo {author}
  {\bibfnamefont {A.}~\bibnamefont {Forchel}}, \ and\ \bibinfo {author}
  {\bibfnamefont {Y.}~\bibnamefont {Yamamoto}},\ }\href@noop {} {\bibfield
  {journal} {\bibinfo  {journal} {New J. Phys.}\ }\textbf {\bibinfo {volume}
  {14}},\ \bibinfo {pages} {065002} (\bibinfo {year} {2012})}\BibitemShut
  {NoStop}%
\bibitem [{\citenamefont {Langbein}\ \emph {et~al.}(1999)\citenamefont
  {Langbein}, \citenamefont {Hvam},\ and\ \citenamefont
  {Zimmermann}}]{Langbein1999}%
  \BibitemOpen
  \bibfield  {author} {\bibinfo {author} {\bibfnamefont {W.}~\bibnamefont
  {Langbein}}, \bibinfo {author} {\bibfnamefont {J.~M.}\ \bibnamefont {Hvam}},
  \ and\ \bibinfo {author} {\bibfnamefont {R.}~\bibnamefont {Zimmermann}},\
  }\href@noop {} {\bibfield  {journal} {\bibinfo  {journal} {Phys. Rev. Lett.}\
  }\textbf {\bibinfo {volume} {82}},\ \bibinfo {pages} {1040} (\bibinfo {year}
  {1999})}\BibitemShut {NoStop}%
\bibitem [{\citenamefont {Wouters}\ and\ \citenamefont
  {Carusotto}(2007)}]{WoutersCarusoto2007}%
  \BibitemOpen
  \bibfield  {author} {\bibinfo {author} {\bibfnamefont {M.}~\bibnamefont
  {Wouters}}\ and\ \bibinfo {author} {\bibfnamefont {I.}~\bibnamefont
  {Carusotto}},\ }\href@noop {} {\bibfield  {journal} {\bibinfo  {journal}
  {Phys. Rev. Lett.}\ }\textbf {\bibinfo {volume} {99}},\ \bibinfo {pages}
  {140402} (\bibinfo {year} {2007})}\BibitemShut {NoStop}%
\bibitem [{\citenamefont {Keeling}\ and\ \citenamefont
  {Berloff}(2008)}]{Keeling2008}%
  \BibitemOpen
  \bibfield  {author} {\bibinfo {author} {\bibfnamefont {J.}~\bibnamefont
  {Keeling}}\ and\ \bibinfo {author} {\bibfnamefont {N.~G.}\ \bibnamefont
  {Berloff}},\ }\href@noop {} {\bibfield  {journal} {\bibinfo  {journal} {Phys.
  Rev. Lett.}\ }\textbf {\bibinfo {volume} {100}},\ \bibinfo {pages} {250401}
  (\bibinfo {year} {2008})}\BibitemShut {NoStop}%
\bibitem [{\citenamefont {Eastham}(2008)}]{Eastham2008}%
  \BibitemOpen
  \bibfield  {author} {\bibinfo {author} {\bibfnamefont {P.~R.}\ \bibnamefont
  {Eastham}},\ }\href@noop {} {\bibfield  {journal} {\bibinfo  {journal} {Phys.
  Rev. B}\ }\textbf {\bibinfo {volume} {78}},\ \bibinfo {pages} {035319}
  (\bibinfo {year} {2008})}\BibitemShut {NoStop}%
\bibitem [{\citenamefont {Rodrigues}\ \emph {et~al.}(2012)\citenamefont
  {Rodrigues}, \citenamefont {Kevrekidis}, \citenamefont {Cuevas},
  \citenamefont {Carretero-Gonzalez},\ and\ \citenamefont
  {Frantzeskakis}}]{Rodrigues2012}%
  \BibitemOpen
  \bibfield  {author} {\bibinfo {author} {\bibfnamefont {A.~S.}\ \bibnamefont
  {Rodrigues}}, \bibinfo {author} {\bibfnamefont {P.~G.}\ \bibnamefont
  {Kevrekidis}}, \bibinfo {author} {\bibfnamefont {J.}~\bibnamefont {Cuevas}},
  \bibinfo {author} {\bibfnamefont {R.}~\bibnamefont {Carretero-Gonzalez}}, \
  and\ \bibinfo {author} {\bibfnamefont {D.~J.}\ \bibnamefont
  {Frantzeskakis}},\ }\href@noop {} {\bibfield  {journal} {\bibinfo  {journal}
  {arXiv:1205.6262}\ } (\bibinfo {year} {2012})}\BibitemShut {NoStop}%
\bibitem [{\citenamefont {{Szyma\'{n}ska}}\ \emph {et~al.}(2007)\citenamefont
  {{Szyma\'{n}ska}}, \citenamefont {Keeling},\ and\ \citenamefont
  {Littlewood}}]{SzymanskaKeelingLittlewood2007}%
  \BibitemOpen
  \bibfield  {author} {\bibinfo {author} {\bibfnamefont {M.~H.}\ \bibnamefont
  {{Szyma\'{n}ska}}}, \bibinfo {author} {\bibfnamefont {J.}~\bibnamefont
  {Keeling}}, \ and\ \bibinfo {author} {\bibfnamefont {P.~B.}\ \bibnamefont
  {Littlewood}},\ }\href@noop {} {\bibfield  {journal} {\bibinfo  {journal}
  {Phys. Rev. B}\ }\textbf {\bibinfo {volume} {75}},\ \bibinfo {pages} {195331}
  (\bibinfo {year} {2007})}\BibitemShut {NoStop}%
\bibitem [{\citenamefont {Tassone}\ and\ \citenamefont
  {Yamamoto}(2000)}]{TassoneYamamoto2000}%
  \BibitemOpen
  \bibfield  {author} {\bibinfo {author} {\bibfnamefont {F.}~\bibnamefont
  {Tassone}}\ and\ \bibinfo {author} {\bibfnamefont {Y.}~\bibnamefont
  {Yamamoto}},\ }\href@noop {} {\bibfield  {journal} {\bibinfo  {journal}
  {Phys. Rev. A}\ }\textbf {\bibinfo {volume} {62}},\ \bibinfo {pages} {063809}
  (\bibinfo {year} {2000})}\BibitemShut {NoStop}%
\bibitem [{\citenamefont {Haug}\ \emph {et~al.}(2012)\citenamefont {Haug},
  \citenamefont {Doan}, \citenamefont {Cao},\ and\ \citenamefont
  {Thoai}}]{HaugThoai2012}%
  \BibitemOpen
  \bibfield  {author} {\bibinfo {author} {\bibfnamefont {H.}~\bibnamefont
  {Haug}}, \bibinfo {author} {\bibfnamefont {T.~D.}\ \bibnamefont {Doan}},
  \bibinfo {author} {\bibfnamefont {H.~T.}\ \bibnamefont {Cao}}, \ and\
  \bibinfo {author} {\bibfnamefont {D.~B.~T.}\ \bibnamefont {Thoai}},\
  }\href@noop {} {\bibfield  {journal} {\bibinfo  {journal} {Phys. Rev. B}\
  }\textbf {\bibinfo {volume} {85}},\ \bibinfo {pages} {205310} (\bibinfo
  {year} {2012})}\BibitemShut {NoStop}%
\bibitem [{\citenamefont {Wouters}\ and\ \citenamefont
  {Savona}(2009)}]{wouters09}%
  \BibitemOpen
  \bibfield  {author} {\bibinfo {author} {\bibfnamefont {M.}~\bibnamefont
  {Wouters}}\ and\ \bibinfo {author} {\bibfnamefont {V.}~\bibnamefont
  {Savona}},\ }\href@noop {} {\bibfield  {journal} {\bibinfo  {journal} {Phys.
  Rev. B}\ }\textbf {\bibinfo {volume} {79}},\ \bibinfo {pages} {165302}
  (\bibinfo {year} {2009})}\BibitemShut {NoStop}%
\bibitem [{\citenamefont {Whittaker}\ and\ \citenamefont
  {Eastham}(2009)}]{Whittaker2009}%
  \BibitemOpen
  \bibfield  {author} {\bibinfo {author} {\bibfnamefont {D.~M.}\ \bibnamefont
  {Whittaker}}\ and\ \bibinfo {author} {\bibfnamefont {P.~R.}\ \bibnamefont
  {Eastham}},\ }\href@noop {} {\bibfield  {journal} {\bibinfo  {journal} {EPL}\
  }\textbf {\bibinfo {volume} {87}},\ \bibinfo {pages} {27002} (\bibinfo {year}
  {2009})}\BibitemShut {NoStop}%
\bibitem [{\citenamefont {Schwendimann}\ \emph {et~al.}(2010)\citenamefont
  {Schwendimann}, \citenamefont {Quattropani},\ and\ \citenamefont
  {Sarchi}}]{SchwendimannQuattropaniSarchi2010}%
  \BibitemOpen
  \bibfield  {author} {\bibinfo {author} {\bibfnamefont {P.}~\bibnamefont
  {Schwendimann}}, \bibinfo {author} {\bibfnamefont {A.}~\bibnamefont
  {Quattropani}}, \ and\ \bibinfo {author} {\bibfnamefont {D.}~\bibnamefont
  {Sarchi}},\ }\href@noop {} {\bibfield  {journal} {\bibinfo  {journal} {Phys.
  Rev. B}\ }\textbf {\bibinfo {volume} {82}},\ \bibinfo {pages} {205329}
  (\bibinfo {year} {2010})}\BibitemShut {NoStop}%
\bibitem [{\citenamefont {Porras}\ and\ \citenamefont
  {Tejedor}(2003)}]{PorrasTejedor2003}%
  \BibitemOpen
  \bibfield  {author} {\bibinfo {author} {\bibfnamefont {D.}~\bibnamefont
  {Porras}}\ and\ \bibinfo {author} {\bibfnamefont {C.}~\bibnamefont
  {Tejedor}},\ }\href@noop {} {\bibfield  {journal} {\bibinfo  {journal} {Phys.
  Rev. B}\ }\textbf {\bibinfo {volume} {67}},\ \bibinfo {pages} {161310}
  (\bibinfo {year} {2003})}\BibitemShut {NoStop}%
\bibitem [{\citenamefont {Laussy}\ \emph {et~al.}(2004)\citenamefont {Laussy},
  \citenamefont {Malpuech}, \citenamefont {Kavokin},\ and\ \citenamefont
  {Bigenwald}}]{LaussyBigenwald2004}%
  \BibitemOpen
  \bibfield  {author} {\bibinfo {author} {\bibfnamefont {F.~P.}\ \bibnamefont
  {Laussy}}, \bibinfo {author} {\bibfnamefont {G.}~\bibnamefont {Malpuech}},
  \bibinfo {author} {\bibfnamefont {A.}~\bibnamefont {Kavokin}}, \ and\
  \bibinfo {author} {\bibfnamefont {P.}~\bibnamefont {Bigenwald}},\ }\href@noop
  {} {\bibfield  {journal} {\bibinfo  {journal} {Phys. Rev. Lett.}\ }\textbf
  {\bibinfo {volume} {93}},\ \bibinfo {pages} {016402} (\bibinfo {year}
  {2004})}\BibitemShut {NoStop}%
\bibitem [{\citenamefont {Verger}\ \emph {et~al.}(2006)\citenamefont {Verger},
  \citenamefont {Ciuti},\ and\ \citenamefont
  {Carusotto}}]{VergerCarusotto2006}%
  \BibitemOpen
  \bibfield  {author} {\bibinfo {author} {\bibfnamefont {A.}~\bibnamefont
  {Verger}}, \bibinfo {author} {\bibfnamefont {C.}~\bibnamefont {Ciuti}}, \
  and\ \bibinfo {author} {\bibfnamefont {I.}~\bibnamefont {Carusotto}},\
  }\href@noop {} {\bibfield  {journal} {\bibinfo  {journal} {Phys. Rev. B}\
  }\textbf {\bibinfo {volume} {73}},\ \bibinfo {pages} {193306} (\bibinfo
  {year} {2006})}\BibitemShut {NoStop}%
\bibitem [{\citenamefont {Schwendimann}\ and\ \citenamefont
  {Quattropani}(2008)}]{SchendimannQuattropani2008}%
  \BibitemOpen
  \bibfield  {author} {\bibinfo {author} {\bibfnamefont {P.}~\bibnamefont
  {Schwendimann}}\ and\ \bibinfo {author} {\bibfnamefont {A.}~\bibnamefont
  {Quattropani}},\ }\href@noop {} {\bibfield  {journal} {\bibinfo  {journal}
  {Phys. Rev. B}\ }\textbf {\bibinfo {volume} {77}},\ \bibinfo {pages} {085317}
  (\bibinfo {year} {2008})}\BibitemShut {NoStop}%
\bibitem [{\citenamefont {Savenko}\ \emph {et~al.}(2011)\citenamefont
  {Savenko}, \citenamefont {Magnusson},\ and\ \citenamefont
  {Shelykh}}]{Shelykh2011}%
  \BibitemOpen
  \bibfield  {author} {\bibinfo {author} {\bibfnamefont {I.~G.}\ \bibnamefont
  {Savenko}}, \bibinfo {author} {\bibfnamefont {E.~B.}\ \bibnamefont
  {Magnusson}}, \ and\ \bibinfo {author} {\bibfnamefont {I.~A.}\ \bibnamefont
  {Shelykh}},\ }\href@noop {} {\bibfield  {journal} {\bibinfo  {journal} {Phys.
  Rev. B}\ }\textbf {\bibinfo {volume} {83}},\ \bibinfo {pages} {165316}
  (\bibinfo {year} {2011})}\BibitemShut {NoStop}%
\bibitem [{\citenamefont {Wouters}(2008)}]{Wouters2008}%
  \BibitemOpen
  \bibfield  {author} {\bibinfo {author} {\bibfnamefont {M.}~\bibnamefont
  {Wouters}},\ }\href@noop {} {\bibfield  {journal} {\bibinfo  {journal} {Phys.
  Rev. B}\ }\textbf {\bibinfo {volume} {77}},\ \bibinfo {pages} {121302}
  (\bibinfo {year} {2008})}\BibitemShut {NoStop}%
\bibitem [{\citenamefont {Singh}\ and\ \citenamefont
  {Zubairy}(1980)}]{SinghZubairy1980}%
  \BibitemOpen
  \bibfield  {author} {\bibinfo {author} {\bibfnamefont {S.}~\bibnamefont
  {Singh}}\ and\ \bibinfo {author} {\bibfnamefont {M.~S.}\ \bibnamefont
  {Zubairy}},\ }\href@noop {} {\bibfield  {journal} {\bibinfo  {journal} {Phys.
  Rev. A}\ }\textbf {\bibinfo {volume} {21}},\ \bibinfo {pages} {281} (\bibinfo
  {year} {1980})}\BibitemShut {NoStop}%
\bibitem [{\citenamefont {Kampen}(2007)}]{VanKampen}%
  \BibitemOpen
  \bibfield  {author} {\bibinfo {author} {\bibfnamefont {N.~V.}\ \bibnamefont
  {Kampen}},\ }\href@noop {} {\emph {\bibinfo {title} {{Stochastic Processes in
  Physics and Chemistry}}}},\ \bibinfo {edition} {3rd}\ ed.\ (\bibinfo
  {publisher} {North Holland, Amsterdam},\ \bibinfo {year} {2007})\BibitemShut {NoStop}%
\bibitem [{\citenamefont {Scully}\ and\ \citenamefont
  {Zubairy}(1997)}]{Scully&Zubairy1997}%
  \BibitemOpen
  \bibfield  {author} {\bibinfo {author} {\bibfnamefont {M.~O.}\ \bibnamefont
  {Scully}}\ and\ \bibinfo {author} {\bibfnamefont {M.~S.}\ \bibnamefont
  {Zubairy}},\ }\href@noop {} {\emph {\bibinfo {title} {{Quantum Optics}}}}\
  (\bibinfo  {publisher} {Cambridge University Press, London},\ \bibinfo {year}
  {1997})\BibitemShut {NoStop}%
\bibitem [{\citenamefont {Aleiner}\ \emph {et~al.}(2012)\citenamefont
  {Aleiner}, \citenamefont {Altshuler},\ and\ \citenamefont
  {Rubo}}]{Aleiner2012}%
  \BibitemOpen
  \bibfield  {author} {\bibinfo {author} {\bibfnamefont {I.~L.}\ \bibnamefont
  {Aleiner}}, \bibinfo {author} {\bibfnamefont {B.~L.}\ \bibnamefont
  {Altshuler}}, \ and\ \bibinfo {author} {\bibfnamefont {Y.~G.}\ \bibnamefont
  {Rubo}},\ }\href@noop {} {\bibfield  {journal} {\bibinfo  {journal} {Phys.
  Rev. B}\ }\textbf {\bibinfo {volume} {85}},\ \bibinfo {pages} {121301}
  (\bibinfo {year} {2012})}\BibitemShut {NoStop}%
\bibitem [{\citenamefont {Deng}\ \emph {et~al.}(2010)\citenamefont {Deng},
  \citenamefont {Haug},\ and\ \citenamefont {Yamamoto}}]{Deng2010}%
  \BibitemOpen
  \bibfield  {author} {\bibinfo {author} {\bibfnamefont {H.}~\bibnamefont
  {Deng}}, \bibinfo {author} {\bibfnamefont {H.}~\bibnamefont {Haug}}, \ and\
  \bibinfo {author} {\bibfnamefont {Y.}~\bibnamefont {Yamamoto}},\ }\href@noop
  {} {\bibfield  {journal} {\bibinfo  {journal} {Rev. Mod. Phys.}\ }\textbf
  {\bibinfo {volume} {82}},\ \bibinfo {pages} {1489} (\bibinfo {year}
  {2010})}\BibitemShut {NoStop}%
\bibitem [{\citenamefont {Wertz}\ \emph {et~al.}(2010)\citenamefont {Wertz},
  \citenamefont {Ferrier}, \citenamefont {Solnyshkov}, \citenamefont {Johne},
  \citenamefont {Sanvitto}, \citenamefont {Lema\^{\i}tre}, \citenamefont
  {Sagnes}, \citenamefont {Grousson}, \citenamefont {Kavokin}, \citenamefont
  {Senellart}, \citenamefont {Malpuech},\ and\ \citenamefont
  {Bloch}}]{Wertz2010}%
  \BibitemOpen
  \bibfield  {author} {\bibinfo {author} {\bibfnamefont {E.}~\bibnamefont
  {Wertz}}, \bibinfo {author} {\bibfnamefont {L.}~\bibnamefont {Ferrier}},
  \bibinfo {author} {\bibfnamefont {D.}~\bibnamefont {Solnyshkov}}, \bibinfo
  {author} {\bibfnamefont {R.}~\bibnamefont {Johne}}, \bibinfo {author}
  {\bibfnamefont {D.}~\bibnamefont {Sanvitto}}, \bibinfo {author}
  {\bibfnamefont {A.}~\bibnamefont {Lema\^{\i}tre}}, \bibinfo {author}
  {\bibfnamefont {I.}~\bibnamefont {Sagnes}}, \bibinfo {author} {\bibfnamefont
  {R.}~\bibnamefont {Grousson}}, \bibinfo {author} {\bibfnamefont {A.~V.}\
  \bibnamefont {Kavokin}}, \bibinfo {author} {\bibfnamefont {P.}~\bibnamefont
  {Senellart}}, \bibinfo {author} {\bibfnamefont {G.}~\bibnamefont {Malpuech}},
  \ and\ \bibinfo {author} {\bibfnamefont {J.}~\bibnamefont {Bloch}},\
  }\href@noop {} {\bibfield  {journal} {\bibinfo  {journal} {Nat. Phys.}\
  }\textbf {\bibinfo {volume} {6}},\ \bibinfo {pages} {860} (\bibinfo {year}
  {2010})}\BibitemShut {NoStop}%
\bibitem [{\citenamefont {Ferrier}\ \emph {et~al.}(2011)\citenamefont
  {Ferrier}, \citenamefont {Wertz}, \citenamefont {Johne}, \citenamefont
  {Solnyshkov}, \citenamefont {Senellart}, \citenamefont {Sagnes},
  \citenamefont {Lema\^{\i}tre}, \citenamefont {Malpuech},\ and\ \citenamefont
  {Bloch}}]{Ferrier2011}%
  \BibitemOpen
  \bibfield  {author} {\bibinfo {author} {\bibfnamefont {L.}~\bibnamefont
  {Ferrier}}, \bibinfo {author} {\bibfnamefont {E.}~\bibnamefont {Wertz}},
  \bibinfo {author} {\bibfnamefont {R.}~\bibnamefont {Johne}}, \bibinfo
  {author} {\bibfnamefont {D.~D.}\ \bibnamefont {Solnyshkov}}, \bibinfo
  {author} {\bibfnamefont {P.}~\bibnamefont {Senellart}}, \bibinfo {author}
  {\bibfnamefont {I.}~\bibnamefont {Sagnes}}, \bibinfo {author} {\bibfnamefont
  {A.}~\bibnamefont {Lema\^{\i}tre}}, \bibinfo {author} {\bibfnamefont
  {G.}~\bibnamefont {Malpuech}}, \ and\ \bibinfo {author} {\bibfnamefont
  {J.}~\bibnamefont {Bloch}},\ }\href@noop {} {\bibfield  {journal} {\bibinfo
  {journal} {Phys. Rev. Lett.}\ }\textbf {\bibinfo {volume} {106}},\ \bibinfo
  {pages} {126401} (\bibinfo {year} {2011})}\BibitemShut {NoStop}%
\bibitem [{\citenamefont {Carmichael}\ and\ \citenamefont
  {Walls}(1974)}]{Walls1974}%
  \BibitemOpen
  \bibfield  {author} {\bibinfo {author} {\bibfnamefont {H.~J.}\ \bibnamefont
  {Carmichael}}\ and\ \bibinfo {author} {\bibfnamefont {D.~F.}\ \bibnamefont
  {Walls}},\ }\href@noop {} {\bibfield  {journal} {\bibinfo  {journal} {Phys.
  Rev. A}\ }\textbf {\bibinfo {volume} {9}},\ \bibinfo {pages} {2686} (\bibinfo
  {year} {1974})}\BibitemShut {NoStop}%
\bibitem [{Not()}]{Note1}%
  \BibitemOpen
  \href@noop {} {}\bibinfo {note} {This can be seen by setting $\alpha
  _1=1,n_c>n_s$ and varying $\alpha _2$. It is also pointed out in
  Ref.~\protect \rev@citealpnum {SinghZubairy1980}. In this context it may be
  helpful to define normalized quantities in terms of the coupling strength to
  one mode, rather than the total coupling, e.g., $n_s^\prime =\gamma
  ^2/(4g_1^2)$.}\BibitemShut {Stop}%
\bibitem [{\citenamefont {Racine}(2013)}]{RacineThesis}%
  \BibitemOpen
  \bibfield  {author} {\bibinfo {author} {\bibfnamefont {D.}~\bibnamefont
  {Racine}},\ }\emph {\bibinfo {title} {Quantum Theory of Multimode
  Exciton-polariton Bose-Einstein Condensation}},\ \href@noop {} {Ph.D.
  thesis},\ \bibinfo  {school} {Trinity College Dublin} (\bibinfo {year}
  {2013})\BibitemShut {NoStop}%
\bibitem [{\citenamefont {Rochat}\ \emph {et~al.}(2000)\citenamefont {Rochat},
  \citenamefont {Ciuti}, \citenamefont {Savona}, \citenamefont {Piermarocchi},
  \citenamefont {Quattropani},\ and\ \citenamefont {Schwendimann}}]{Ciuti2000}%
  \BibitemOpen
  \bibfield  {author} {\bibinfo {author} {\bibfnamefont {G.}~\bibnamefont
  {Rochat}}, \bibinfo {author} {\bibfnamefont {C.}~\bibnamefont {Ciuti}},
  \bibinfo {author} {\bibfnamefont {V.}~\bibnamefont {Savona}}, \bibinfo
  {author} {\bibfnamefont {C.}~\bibnamefont {Piermarocchi}}, \bibinfo {author}
  {\bibfnamefont {A.}~\bibnamefont {Quattropani}}, \ and\ \bibinfo {author}
  {\bibfnamefont {P.}~\bibnamefont {Schwendimann}},\ }\href@noop {} {\bibfield
  {journal} {\bibinfo  {journal} {Phys. Rev. B}\ }\textbf {\bibinfo {volume}
  {61}},\ \bibinfo {pages} {13856} (\bibinfo {year} {2000})}\BibitemShut
  {NoStop}%
\bibitem [{\citenamefont {Loudon}(2000)}]{loudonspectrum}%
  \BibitemOpen
  \bibfield  {author} {\bibinfo {author} {\bibfnamefont {R.}~\bibnamefont
  {Loudon}},\ }\enquote {\bibinfo {title} {The quantum theory of light},}\ \
  (\bibinfo  {publisher} {Oxford University Press, London},\ \bibinfo {year} {2000})\
  p.\ \bibinfo {pages} {102},\ \bibinfo {edition} {3rd}\ ed.\BibitemShut
  {Stop}%
\bibitem [{\citenamefont {Swain}(1981)}]{Swain1981}%
  \BibitemOpen
  \bibfield  {author} {\bibinfo {author} {\bibfnamefont {S.}~\bibnamefont
  {Swain}},\ }\href@noop {} {\bibfield  {journal} {\bibinfo  {journal} {J.
  Phys. A: Math. Gen.}\ }\textbf {\bibinfo {volume} {14}},\ \bibinfo {pages}
  {2577} (\bibinfo {year} {1981})}\BibitemShut {NoStop}%
\bibitem [{\citenamefont {Ford}\ and\ \citenamefont
  {O'Connell}(1996)}]{Ford1996}%
  \BibitemOpen
  \bibfield  {author} {\bibinfo {author} {\bibfnamefont {G.~W.}\ \bibnamefont
  {Ford}}\ and\ \bibinfo {author} {\bibfnamefont {R.~F.}\ \bibnamefont
  {O'Connell}},\ }\href@noop {} {\bibfield  {journal} {\bibinfo  {journal}
  {Phys. Rev. Lett.}\ }\textbf {\bibinfo {volume} {77}},\ \bibinfo {pages}
  {798} (\bibinfo {year} {1996})}\BibitemShut {NoStop}%
\bibitem [{\citenamefont {Hamm}(2005)}]{Hamm2005}%
  \BibitemOpen
  \bibfield  {author} {\bibinfo {author} {\bibfnamefont {P.}~\bibnamefont
  {Hamm}},\ }\href@noop {} {\enquote {\bibinfo {title} {Principles of nonlinear
  spectroscopy: a practical approach},}\ }\bibinfo {howpublished} {Lectures of
  the Virtual European University on Lasers} (\bibinfo {year} {2005}),\
  \bibinfo {note} {(unpublished)}\BibitemShut {NoStop}%
\bibitem [{\citenamefont {Kubo}(1954)}]{Kubo1954}%
  \BibitemOpen
  \bibfield  {author} {\bibinfo {author} {\bibfnamefont {R.}~\bibnamefont
  {Kubo}},\ }\href@noop {} {\bibfield  {journal} {\bibinfo  {journal} {J. Phys.
  Soc. Jpn.}\ }\textbf {\bibinfo {volume} {9}},\ \bibinfo {pages} {935}
  (\bibinfo {year} {1954})}\BibitemShut {NoStop}%
\bibitem [{\citenamefont {Leggett}(2001)}]{Leggett2001}%
  \BibitemOpen
  \bibfield  {author} {\bibinfo {author} {\bibfnamefont {A.~J.}\ \bibnamefont
  {Leggett}},\ }\href@noop {} {\bibfield  {journal} {\bibinfo  {journal} {Rev.
  Mod. Phys.}\ }\textbf {\bibinfo {volume} {73}},\ \bibinfo {pages} {307}
  (\bibinfo {year} {2001})}\BibitemShut {NoStop}%
\bibitem [{\citenamefont {Raftery}\ \emph {et~al.}(2013)\citenamefont
  {Raftery}, \citenamefont {Sadri}, \citenamefont {Schmidt}, \citenamefont
  {T\"ureci},\ and\ \citenamefont {Houck}}]{Raftery2014}%
  \BibitemOpen
  \bibfield  {author} {\bibinfo {author} {\bibfnamefont {J.}~\bibnamefont
  {Raftery}}, \bibinfo {author} {\bibfnamefont {D.}~\bibnamefont {Sadri}},
  \bibinfo {author} {\bibfnamefont {S.}~\bibnamefont {Schmidt}}, \bibinfo
  {author} {\bibfnamefont {H.~E.}\ \bibnamefont {T\"ureci}}, \ and\ \bibinfo
  {author} {\bibfnamefont {A.~A.}\ \bibnamefont {Houck}},\ }\href@noop {}
  {\bibfield  {journal} {\bibinfo  {journal} {arXiv:1312.2963}\ } (\bibinfo
  {year} {2013})}\BibitemShut {NoStop}%
\bibitem [{\citenamefont {Sarchi}\ \emph {et~al.}(2008)\citenamefont {Sarchi},
  \citenamefont {Carusotto}, \citenamefont {Wouters},\ and\ \citenamefont
  {Savona}}]{sarchi2008}%
  \BibitemOpen
  \bibfield  {author} {\bibinfo {author} {\bibfnamefont {D.}~\bibnamefont
  {Sarchi}}, \bibinfo {author} {\bibfnamefont {I.}~\bibnamefont {Carusotto}},
  \bibinfo {author} {\bibfnamefont {M.}~\bibnamefont {Wouters}}, \ and\
  \bibinfo {author} {\bibfnamefont {V.}~\bibnamefont {Savona}},\ }\href@noop {}
  {\bibfield  {journal} {\bibinfo  {journal} {Phys. Rev. B}\ }\textbf {\bibinfo
  {volume} {77}},\ \bibinfo {pages} {125324} (\bibinfo {year}
  {2008})}\BibitemShut {NoStop}%
\bibitem [{\citenamefont {Magnusson}\ \emph {et~al.}(2010)\citenamefont
  {Magnusson}, \citenamefont {Flayac}, \citenamefont {Malpuech},\ and\
  \citenamefont {Shelykh}}]{MagnussonShelykh2010}%
  \BibitemOpen
  \bibfield  {author} {\bibinfo {author} {\bibfnamefont {E.~B.}\ \bibnamefont
  {Magnusson}}, \bibinfo {author} {\bibfnamefont {H.}~\bibnamefont {Flayac}},
  \bibinfo {author} {\bibfnamefont {G.}~\bibnamefont {Malpuech}}, \ and\
  \bibinfo {author} {\bibfnamefont {I.~A.}\ \bibnamefont {Shelykh}},\
  }\href@noop {} {\bibfield  {journal} {\bibinfo  {journal} {Phys. Rev. B}\
  }\textbf {\bibinfo {volume} {82}},\ \bibinfo {pages} {195312} (\bibinfo
  {year} {2010})}\BibitemShut {NoStop}%
\bibitem [{\citenamefont {Read}\ \emph {et~al.}(2010)\citenamefont {Read},
  \citenamefont {Rubo},\ and\ \citenamefont {Kavokin}}]{read2010}%
  \BibitemOpen
  \bibfield  {author} {\bibinfo {author} {\bibfnamefont {D.}~\bibnamefont
  {Read}}, \bibinfo {author} {\bibfnamefont {Y.~G.}\ \bibnamefont {Rubo}}, \
  and\ \bibinfo {author} {\bibfnamefont {A.~V.}\ \bibnamefont {Kavokin}},\
  }\href@noop {} {\bibfield  {journal} {\bibinfo  {journal} {Phys. Rev. B}\
  }\textbf {\bibinfo {volume} {81}},\ \bibinfo {pages} {235315} (\bibinfo
  {year} {2010})}\BibitemShut {NoStop}%
\bibitem [{\citenamefont {Eastham}\ and\ \citenamefont
  {Littlewood}(2006)}]{eastham2006}%
  \BibitemOpen
  \bibfield  {author} {\bibinfo {author} {\bibfnamefont {P.~R.}\ \bibnamefont
  {Eastham}}\ and\ \bibinfo {author} {\bibfnamefont {P.~B.}\ \bibnamefont
  {Littlewood}},\ }\href@noop {} {\bibfield  {journal} {\bibinfo  {journal}
  {Phys. Rev. B}\ }\textbf {\bibinfo {volume} {73}},\ \bibinfo {pages} {085306}
  (\bibinfo {year} {2006})}\BibitemShut {NoStop}%
\bibitem [{\citenamefont {Janot}\ \emph {et~al.}(2013)\citenamefont {Janot},
  \citenamefont {Hyart}, \citenamefont {Eastham},\ and\ \citenamefont
  {Rosenow}}]{Janot:2013}%
  \BibitemOpen
  \bibfield  {author} {\bibinfo {author} {\bibfnamefont {A.}~\bibnamefont
  {Janot}}, \bibinfo {author} {\bibfnamefont {T.}~\bibnamefont {Hyart}},
  \bibinfo {author} {\bibfnamefont {P.~R.}\ \bibnamefont {Eastham}}, \ and\
  \bibinfo {author} {\bibfnamefont {B.}~\bibnamefont {Rosenow}},\ }\href@noop
  {} {\bibfield  {journal} {\bibinfo  {journal} {Phys. Rev. Lett.}\ }\textbf
  {\bibinfo {volume} {111}},\ \bibinfo {pages} {230403} (\bibinfo {year}
  {2013})}\BibitemShut {NoStop}%
\bibitem [{\citenamefont {Malpuech}\ \emph {et~al.}(2007)\citenamefont
  {Malpuech}, \citenamefont {Solnyshkov}, \citenamefont {Ouerdane},
  \citenamefont {Glazov},\ and\ \citenamefont {Shelykh}}]{Malpuech:2007}%
  \BibitemOpen
  \bibfield  {author} {\bibinfo {author} {\bibfnamefont {G.}~\bibnamefont
  {Malpuech}}, \bibinfo {author} {\bibfnamefont {D.~D.}\ \bibnamefont
  {Solnyshkov}}, \bibinfo {author} {\bibfnamefont {H.}~\bibnamefont
  {Ouerdane}}, \bibinfo {author} {\bibfnamefont {M.~M.}\ \bibnamefont
  {Glazov}}, \ and\ \bibinfo {author} {\bibfnamefont {I.}~\bibnamefont
  {Shelykh}},\ }\href@noop {} {\bibfield  {journal} {\bibinfo  {journal} {Phys.
  Rev. Lett.}\ }\textbf {\bibinfo {volume} {98}},\ \bibinfo {pages} {206402}
  (\bibinfo {year} {2007})}\BibitemShut {NoStop}%
\bibitem [{\citenamefont {Carusotto}\ and\ \citenamefont
  {Ciuti}(2013)}]{carusottrev:2013}%
  \BibitemOpen
  \bibfield  {author} {\bibinfo {author} {\bibfnamefont {I.}~\bibnamefont
  {Carusotto}}\ and\ \bibinfo {author} {\bibfnamefont {C.}~\bibnamefont
  {Ciuti}},\ }\href@noop {} {\bibfield  {journal} {\bibinfo  {journal} {Rev.
  Mod. Phys.}\ }\textbf {\bibinfo {volume} {85}},\ \bibinfo {pages} {299}
  (\bibinfo {year} {2013})}\BibitemShut {NoStop}%
\bibitem [{\citenamefont {Wouters}\ \emph {et~al.}(2008)\citenamefont
  {Wouters}, \citenamefont {Carusotto},\ and\ \citenamefont
  {Ciuti}}]{Wouters2:2008}%
  \BibitemOpen
  \bibfield  {author} {\bibinfo {author} {\bibfnamefont {M.}~\bibnamefont
  {Wouters}}, \bibinfo {author} {\bibfnamefont {I.}~\bibnamefont {Carusotto}},
  \ and\ \bibinfo {author} {\bibfnamefont {C.}~\bibnamefont {Ciuti}},\
  }\href@noop {} {\bibfield  {journal} {\bibinfo  {journal} {Phys. Rev. B}\
  }\textbf {\bibinfo {volume} {77}},\ \bibinfo {pages} {115340} (\bibinfo
  {year} {2008})}\BibitemShut {NoStop}%
\bibitem [{\citenamefont {Poddubny}\ \emph {et~al.}(2013)\citenamefont
  {Poddubny}, \citenamefont {Glazov},\ and\ \citenamefont
  {N.S.}}]{Poddubny2013}%
  \BibitemOpen
  \bibfield  {author} {\bibinfo {author} {\bibfnamefont {A.}~\bibnamefont
  {Poddubny}}, \bibinfo {author} {\bibfnamefont {M.}~\bibnamefont {Glazov}}, \
  and\ \bibinfo {author} {\bibfnamefont {A.}~\bibnamefont {N.S.}},\ }\href@noop
  {} {\bibfield  {journal} {\bibinfo  {journal} {New J. Phys.}\ }\textbf
  {\bibinfo {volume} {15}},\ \bibinfo {pages} {025016} (\bibinfo {year}
  {2013})}\BibitemShut {NoStop}%
\bibitem [{\citenamefont {Shelykh}\ \emph {et~al.}(2010)\citenamefont
  {Shelykh}, \citenamefont {Kavokin}, \citenamefont {Rubo}, \citenamefont
  {Liew},\ and\ \citenamefont {Malpuech}}]{ShelykhMalpuech2010}%
  \BibitemOpen
  \bibfield  {author} {\bibinfo {author} {\bibfnamefont {I.~A.}\ \bibnamefont
  {Shelykh}}, \bibinfo {author} {\bibfnamefont {A.~V.}\ \bibnamefont
  {Kavokin}}, \bibinfo {author} {\bibfnamefont {Y.}~\bibnamefont {Rubo}},
  \bibinfo {author} {\bibfnamefont {T.}~\bibnamefont {Liew}}, \ and\ \bibinfo
  {author} {\bibfnamefont {G.}~\bibnamefont {Malpuech}},\ }\href@noop {}
  {\bibfield  {journal} {\bibinfo  {journal} {Semicond. Sci. Technol.}\
  }\textbf {\bibinfo {volume} {25}},\ \bibinfo {pages} {013001} (\bibinfo
  {year} {2010})}\BibitemShut {NoStop}%
\bibitem [{\citenamefont {Kasprzak}\ \emph {et~al.}(2007)\citenamefont
  {Kasprzak}, \citenamefont {Andr\'e}, \citenamefont {Dang}, \citenamefont
  {Shelykh}, \citenamefont {Kavokin}, \citenamefont {Rubo}, \citenamefont
  {Kavokin},\ and\ \citenamefont {Malpuech}}]{Kasprzak2007}%
  \BibitemOpen
  \bibfield  {author} {\bibinfo {author} {\bibfnamefont {J.}~\bibnamefont
  {Kasprzak}}, \bibinfo {author} {\bibfnamefont {R.}~\bibnamefont {Andr\'e}},
  \bibinfo {author} {\bibfnamefont {L.~S.}\ \bibnamefont {Dang}}, \bibinfo
  {author} {\bibfnamefont {I.~A.}\ \bibnamefont {Shelykh}}, \bibinfo {author}
  {\bibfnamefont {A.~V.}\ \bibnamefont {Kavokin}}, \bibinfo {author}
  {\bibfnamefont {Y.~G.}\ \bibnamefont {Rubo}}, \bibinfo {author}
  {\bibfnamefont {K.~V.}\ \bibnamefont {Kavokin}}, \ and\ \bibinfo {author}
  {\bibfnamefont {G.}~\bibnamefont {Malpuech}},\ }\href@noop {} {\bibfield
  {journal} {\bibinfo  {journal} {Phys. Rev. B}\ }\textbf {\bibinfo {volume}
  {75}},\ \bibinfo {pages} {045326} (\bibinfo {year} {2007})}\BibitemShut
  {NoStop}%
\bibitem [{\citenamefont {Klopotowski}\ \emph {et~al.}(2006)\citenamefont
  {Klopotowski}, \citenamefont {Martin}, \citenamefont {Amo}, \citenamefont
  {Vina}, \citenamefont {Shelykh}, \citenamefont {Glazov}, \citenamefont
  {Malpuech}, \citenamefont {Kavokin},\ and\ \citenamefont
  {André}}]{Klopotowski2006}%
  \BibitemOpen
  \bibfield  {author} {\bibinfo {author} {\bibfnamefont {L.}~\bibnamefont
  {Klopotowski}}, \bibinfo {author} {\bibfnamefont {M.}~\bibnamefont {Martin}},
  \bibinfo {author} {\bibfnamefont {A.}~\bibnamefont {Amo}}, \bibinfo {author}
  {\bibfnamefont {L.}~\bibnamefont {Vina}}, \bibinfo {author} {\bibfnamefont
  {I.}~\bibnamefont {Shelykh}}, \bibinfo {author} {\bibfnamefont
  {M.}~\bibnamefont {Glazov}}, \bibinfo {author} {\bibfnamefont
  {G.}~\bibnamefont {Malpuech}}, \bibinfo {author} {\bibfnamefont
  {A.}~\bibnamefont {Kavokin}}, \ and\ \bibinfo {author} {\bibfnamefont
  {R.}~\bibnamefont {Andr{\'e}}},\ }\href@noop {} {\bibfield  {journal} {\bibinfo
  {journal} {Solid State Comm.}\ }\textbf {\bibinfo {volume} {139}},\ \bibinfo
  {pages} {511} (\bibinfo {year} {2006})}\BibitemShut {NoStop}%
\bibitem [{\citenamefont {Shelykh}\ \emph {et~al.}(2006)\citenamefont
  {Shelykh}, \citenamefont {Rubo}, \citenamefont {Malpuech}, \citenamefont
  {Solnyshkov},\ and\ \citenamefont {Kavokin}}]{Shelykh2006}%
  \BibitemOpen
  \bibfield  {author} {\bibinfo {author} {\bibfnamefont {I.~A.}\ \bibnamefont
  {Shelykh}}, \bibinfo {author} {\bibfnamefont {Y.~G.}\ \bibnamefont {Rubo}},
  \bibinfo {author} {\bibfnamefont {G.}~\bibnamefont {Malpuech}}, \bibinfo
  {author} {\bibfnamefont {D.~D.}\ \bibnamefont {Solnyshkov}}, \ and\ \bibinfo
  {author} {\bibfnamefont {A.}~\bibnamefont {Kavokin}},\ }\href@noop {}
  {\bibfield  {journal} {\bibinfo  {journal} {Phys. Rev. Lett.}\ }\textbf
  {\bibinfo {volume} {97}},\ \bibinfo {pages} {066402} (\bibinfo {year}
  {2006})}\BibitemShut {NoStop}%
\bibitem [{\citenamefont {Laussy}\ \emph {et~al.}(2006)\citenamefont {Laussy},
  \citenamefont {Shelykh}, \citenamefont {Malpuech},\ and\ \citenamefont
  {Kavokin}}]{Laussy2006}%
  \BibitemOpen
  \bibfield  {author} {\bibinfo {author} {\bibfnamefont {F.~P.}\ \bibnamefont
  {Laussy}}, \bibinfo {author} {\bibfnamefont {I.~A.}\ \bibnamefont {Shelykh}},
  \bibinfo {author} {\bibfnamefont {G.}~\bibnamefont {Malpuech}}, \ and\
  \bibinfo {author} {\bibfnamefont {A.}~\bibnamefont {Kavokin}},\ }\href@noop
  {} {\bibfield  {journal} {\bibinfo  {journal} {Phys. Rev. B}\ }\textbf
  {\bibinfo {volume} {73}},\ \bibinfo {pages} {035315} (\bibinfo {year}
  {2006})}\BibitemShut {NoStop}%
\bibitem [{\citenamefont {Krizhanovskii}\ \emph {et~al.}(2009)\citenamefont
  {Krizhanovskii}, \citenamefont {Lagoudakis}, \citenamefont {Wouters},
  \citenamefont {Pietka}, \citenamefont {Bradley}, \citenamefont {Guda},
  \citenamefont {Whittaker}, \citenamefont {Skolnick}, \citenamefont
  {Deveaud-Pl\'edran}, \citenamefont {Richard}, \citenamefont {Andr\'e},\ and\
  \citenamefont {Dang}}]{Krizhanovskii2009}%
  \BibitemOpen
  \bibfield  {author} {\bibinfo {author} {\bibfnamefont {D.~N.}\ \bibnamefont
  {Krizhanovskii}}, \bibinfo {author} {\bibfnamefont {K.~G.}\ \bibnamefont
  {Lagoudakis}}, \bibinfo {author} {\bibfnamefont {M.}~\bibnamefont {Wouters}},
  \bibinfo {author} {\bibfnamefont {B.}~\bibnamefont {Pietka}}, \bibinfo
  {author} {\bibfnamefont {R.~A.}\ \bibnamefont {Bradley}}, \bibinfo {author}
  {\bibfnamefont {K.}~\bibnamefont {Guda}}, \bibinfo {author} {\bibfnamefont
  {D.~M.}\ \bibnamefont {Whittaker}}, \bibinfo {author} {\bibfnamefont {M.~S.}\
  \bibnamefont {Skolnick}}, \bibinfo {author} {\bibfnamefont {B.}~\bibnamefont
  {Deveaud-Pl\'edran}}, \bibinfo {author} {\bibfnamefont {M.}~\bibnamefont
  {Richard}}, \bibinfo {author} {\bibfnamefont {R.}~\bibnamefont {Andr\'e}}, \
  and\ \bibinfo {author} {\bibfnamefont {L.~S.}\ \bibnamefont {Dang}},\
  }\href@noop {} {\bibfield  {journal} {\bibinfo  {journal} {Phys. Rev. B}\
  }\textbf {\bibinfo {volume} {80}},\ \bibinfo {pages} {045317} (\bibinfo
  {year} {2009})}\BibitemShut {NoStop}%
\bibitem [{\citenamefont {Mart{\'i}n}\ \emph {et~al.}(2002)\citenamefont
  {Mart{\'i}n}, \citenamefont {Aichmayr}, \citenamefont {Vi{\~n}a},\ and\
  \citenamefont {Andr{\'e}}}]{Martin2002}%
  \BibitemOpen
  \bibfield  {author} {\bibinfo {author} {\bibfnamefont {M.~D.}\ \bibnamefont
  {Mart{\'i}n}}, \bibinfo {author} {\bibfnamefont {G.}~\bibnamefont
  {Aichmayr}}, \bibinfo {author} {\bibfnamefont {L.}~\bibnamefont {Vi{\~n}a}},
  \ and\ \bibinfo {author} {\bibfnamefont {R.}~\bibnamefont {Andr{\'e}}},\
  }\href@noop {} {\bibfield  {journal} {\bibinfo  {journal} {Phys. Rev. Lett.}\
  }\textbf {\bibinfo {volume} {89}},\ \bibinfo {pages} {077402} (\bibinfo
  {year} {2002})}\BibitemShut {NoStop}%
\bibitem [{\citenamefont {Glazov}\ \emph {et~al.}(2013)\citenamefont {Glazov},
  \citenamefont {Semina}, \citenamefont {Sherman},\ and\ \citenamefont
  {Kavokin}}]{GlazovKavokin2013}%
  \BibitemOpen
  \bibfield  {author} {\bibinfo {author} {\bibfnamefont {M.~M.}\ \bibnamefont
  {Glazov}}, \bibinfo {author} {\bibfnamefont {M.~A.}\ \bibnamefont {Semina}},
  \bibinfo {author} {\bibfnamefont {E.~Y.}\ \bibnamefont {Sherman}}, \ and\
  \bibinfo {author} {\bibfnamefont {A.~V.}\ \bibnamefont {Kavokin}},\
  }\href@noop {} {\bibfield  {journal} {\bibinfo  {journal} {Phys. Rev. B}\
  }\textbf {\bibinfo {volume} {88}},\ \bibinfo {pages} {041309} (\bibinfo
  {year} {2013})}\BibitemShut {NoStop}%
\bibitem [{\citenamefont {Sabatier}\ \emph {et~al.}(2007)\citenamefont
  {Sabatier}, \citenamefont {Agrawal},\ and\ \citenamefont
  {Machado}}]{FracCal2007}%
  \BibitemOpen
  \bibfield  {author} {\bibinfo {author} {\bibfnamefont {J.}~\bibnamefont
  {Sabatier}}, \bibinfo {author} {\bibfnamefont {O.}~\bibnamefont {Agrawal}}, \
  and\ \bibinfo {author} {\bibfnamefont {J.~T.}\ \bibnamefont {Machado}},\
  }\href@noop {} {\emph {\bibinfo {title} {Advances in Fractional Calculus:
  Theoretical Developments and Applications in Physics and Engineering}}}\
  (\bibinfo  {publisher} {Springer, Berlin},\ \bibinfo {year} {2007})\BibitemShut
  {NoStop}%
\bibitem [{\citenamefont {Das}(2011)}]{FracCal2011}%
  \BibitemOpen
  \bibfield  {author} {\bibinfo {author} {\bibfnamefont {S.}~\bibnamefont
  {Das}},\ }\href@noop {} {\emph {\bibinfo {title} {Functional Fractional
  Calculus}}}\ (\bibinfo  {publisher} {Springer, Berlin},\ \bibinfo {year}
  {2011})\BibitemShut {NoStop}%
\bibitem [{\citenamefont {Javanainen}(1986)}]{javanainen1986}%
  \BibitemOpen
  \bibfield  {author} {\bibinfo {author} {\bibfnamefont {J.}~\bibnamefont
  {Javanainen}},\ }\href@noop {} {\bibfield  {journal} {\bibinfo  {journal}
  {Phys. Rev. Lett.}\ }\textbf {\bibinfo {volume} {57}},\ \bibinfo {pages}
  {3164} (\bibinfo {year} {1986})}\BibitemShut {NoStop}%
\bibitem [{\citenamefont {Raghavan}\ \emph {et~al.}(1999)\citenamefont
  {Raghavan}, \citenamefont {Smerzi}, \citenamefont {Fantoni},\ and\
  \citenamefont {Shenoy}}]{BECJosephTheory}%
  \BibitemOpen
  \bibfield  {author} {\bibinfo {author} {\bibfnamefont {S.}~\bibnamefont
  {Raghavan}}, \bibinfo {author} {\bibfnamefont {A.}~\bibnamefont {Smerzi}},
  \bibinfo {author} {\bibfnamefont {S.}~\bibnamefont {Fantoni}}, \ and\
  \bibinfo {author} {\bibfnamefont {S.~R.}\ \bibnamefont {Shenoy}},\
  }\href@noop {} {\bibfield  {journal} {\bibinfo  {journal} {Phys. Rev. A}\
  }\textbf {\bibinfo {volume} {59}},\ \bibinfo {pages} {620} (\bibinfo {year}
  {1999})}\BibitemShut {NoStop}%
\end{thebibliography}
\end{document}